\UseRawInputEncoding 

\documentclass{nature3}
\usepackage{graphicx}
\usepackage{xcolor}
\usepackage{float}
\usepackage{longtable}
\usepackage{tabularx}
\usepackage{url}
\usepackage{amsmath}
\usepackage{amssymb}
\usepackage{gensymb}
\usepackage{subcaption}
\usepackage{courier}
\usepackage{adjustbox}
\usepackage{rotating}
\usepackage{inputenc}
\usepackage{subcaption}
\usepackage{graphbox}
\usepackage{tablefootnote}

\newbox{\bigpicturebox}

\usepackage[labelfont=bf]{caption}

\usepackage[super]{natbib}

\usepackage[colorlinks=true,citecolor=blue,urlcolor=cyan]{hyperref}
\usepackage{astjnlabbrev}

\usepackage[affil-it]{authblk}

\makeatletter
\def\@maketitle{%
  \newpage
  \begin{center}%
  \let \footnote \thanks
    {\Large \@title \par}%
    \vskip -3em
    {\large
      \begin{tabular}[t]{c}%
        \@author
      \end{tabular}\par}%
  \end{center}%
  \par
  \vskip 0.5em}
\makeatother

\title{\large{Diversity in the haziness and chemistry of temperate sub-Neptunes}}

\author[1]{Pierre-Alexis Roy} 
\affil[1]{Department of Physics and Trottier Institute for Research on Exoplanets, Universit\'{e} de Montr\'{e}al, Montreal, QC, Canada}

\author[1,2]{Bj\"{o}rn Benneke} 
\affil[2]{Department of Earth, Planetary, and Space Sciences, University of California, Los Angeles, CA 90095, USA}

\author[1,3]{Marylou Fournier-Tondreau}
\affil[3]{Department of Physics, University of Oxford, Parks Rd, Oxford, OX1 3PU, UK}

\author[1]{Louis-Philippe Coulombe}

\author[1,4]{Caroline Piaulet-Ghorayeb}
\affil[4]{Department of Astronomy \& Astrophysics, University of Chicago, 5640 South Ellis Avenue, Chicago, IL 60637, USA}

\author[1]{David Lafreni\`ere}

\author[1]{Romain Allart}

\author[5,6]{Nicolas B.\ Cowan}
\affil[5]{Department of Physics, McGill University, 3600 University St, Montreal, QC H3A 2T8, Canada}
\affil[6]{Department of Earth \& Planetary Sciences, McGill University, 3450 University St, Montréal, H3A 2A7, Canada}

\author[1,7]{Lisa Dang}
\affil[7]{Department of Physics and Astronomy and Waterloo Centre for Astrophysics, University of Waterloo, Waterloo, ON N2L 3G1, Canada}

\author[8,9]{Doug Johnstone}
\affil[8]{NRC Herzberg Astronomy and Astrophysics, 5071 West Saanich Rd, Victoria, BC, V9E 2E7, Canada }
\affil[9]{Department of Physics and Astronomy, University of Victoria, Victoria, BC, V8P 5C2, Canada}

\author[10,11]{Adam B. Langeveld}
\affil[10]{Department of Physics and Astronomy, Johns Hopkins University, 3400 N. Charles Street, Baltimore, MD 21218, USA}
\affil[11]{Department of Astronomy and Carl Sagan Institute, Cornell University, Ithaca, NY 14850, USA}

\author[1,12]{Stefan Pelletier}
\affil[12]{Observatoire astronomique de l'Universit\'e de Gen\`eve, 51 chemin Pegasi 1290 Versoix, Switzerland}

\author[1,4]{Michael Radica}

\author[3]{Jake Taylor}

\author[1]{Lo\"ic Albert}

\author[1]{Ren\'e Doyon}

\author[10]{Laura Flagg}

\author[10]{Ray Jayawardhana}

\author[13,14]{Ryan J. MacDonald}
\affil[13]{Department of Astronomy, University of Michigan, Ann Arbor, MI, USA}
\affil[14]{School of Physics and Astronomy, University of St Andrews, North Haugh, St Andrews, UK}

\author[11]{Jake D. Turner}

\begin{document}
\maketitle

\begin{abstract}

Recent JWST transit observations of the temperate (T$_\mathrm{eq}$=250--400\,K) sub-Neptunes K2-18\,b and TOI-270\,d have revealed strong molecular absorption signatures, corroborating the picture that sub-Neptunes below 400\,K host upper atmospheres mostly free of aerosols. They have also revealed the presence of CO$_2$ which is not expected for cool Neptune-like atmospheres in chemical equilibrium. The presence of this oxidized form of carbon was interpreted as a result of vertical mixing or atmosphere-interior interactions, and potentially points to formation near or past the water ice-line. Here, we present the JWST NIRSpec/PRISM 0.7--5.2\,$\mu$m transmission spectrum of the temperate sub-Neptune LP~791-18\,c (T$_\mathrm{eq}$=355\,K). Despite the planet's similar bulk properties and its equilibrium temperature in between that of K2-18\,b (T$_\mathrm{eq}$=255\,K) and TOI-270\,d (T$_\mathrm{eq}$=384\,K), we find a highly different transmission spectrum for LP~791-18\,c. We reveal the presence of methane (detected at 4.5\,$\sigma$) in the atmosphere of LP~791-18\,c, similarly to TOI-270\,d and K2-18\,b; however, the overall spectrum is dominated by a strong Rayleigh scattering slope due to opaque hazes (detected at 5.4\,$\sigma$) and there is no contribution of CO$_2$ opacity. Overall, we infer a deep metal-enriched atmosphere (250--400$\times$solar), with no signs of disequilibrium chemistry, and with more CH$_4$ than CO$_2$ (CH$_4$/CO$_2$ $>12$ at 2$\sigma$). This CH$_4$/CO$_2$ measurement allows us to infer a bulk composition that is richer in H$_2$ than in H$_2$O for LP~791-18\,c when compared to  K2-18\,b and TOI-270\,d, pointing at a formation inside the water ice-line. The hazy nature of the atmosphere of LP~791-18\,c is also in strong contrast with the hypothesis that temperate sub-Neptunes generally have clear upper atmospheres. This discovery demonstrates that there is intrinsic diversity in sub-Neptune envelopes, as even planets that are analogue in density and temperature show clear differences, both in terms of atmospheric chemistry and aerosol formation, possibly stemming from different formation histories.

\end{abstract}

\pagebreak

\textit{This preprint is the original submitted version of the manuscript and has not undergone peer review (when applicable) or any post-submission improvements or corrections. The Version of Record of this article is published in Nature Astronomy, and is available online at https://www.nature.com/articles/s41550-025-02723-3.}

We observed a primary transit of the temperate sub-Neptune exoplanet LP~791-18\,c \citep{crossfield_super-earth_2019} on 17 June 2023 with the Near Infrared Spectrograph (NIRSpec \citep{jakobsen_near-infrared_2022, birkmann_near-infrared_2022}) onboard the James Webb Space Telescope (JWST) as part of GTO 1201 (P.I.: D.L.). With its 2.438 $\pm$ 0.096 R$_\oplus$ radius and 7.1 $\pm$ 0.7 M$_\oplus$ mass \citep{peterson_temperate_2023}, LP~791-18\,c has a bulk density of 2.69 g/cm$^3$, implying the presence of an atmosphere amenable to characterization through transit spectroscopy. Furthermore, with an equilibrium temperature of 324~K (assuming uniform heat redistribution and a Bond albedo of 0.3 \citep{peterson_temperate_2023}), the sub-Neptune is found in a temperate regime, and has physical properties similar to that of K2-18\,b (R$_\mathrm{p}$ = 2.61 $\pm$ 0.09\,R$_\oplus$, M$_\mathrm{p}$ = 8.63 $\pm$ 1.35\,M$_\oplus$, T$_\mathrm{eq, A_B=0.3}$ = 255 $\pm$ 4\,K) \citep{montet_stellar_2015, cloutier_characterization_2017, benneke_spitzer_2017, benneke_water_2019, madhusudhan_carbon-bearing_2023}. In line with the multiple recent studies pertaining to the nature of the low-temperature sub-Neptunes K2-18\,b and TOI-270\,d \citep{madhusudhan_carbon-bearing_2023, wogan_jwst_2024, shorttle_distinguishing_2024, benneke_jwst_2024}, LP~791-18\,c could host multiple atmospheric compositions, be it a deep miscible atmosphere with efficient vertical mixing \citep{yu_how_2021, benneke_jwst_2024}, or a thin atmosphere atop a water ocean \citep{hu_unveiling_2021, madhusudhan_habitability_2021} or a magma layer \citep{kite_atmosphere_2020, shorttle_distinguishing_2024}. Transmission spectroscopy can shed valuable insights into its chemical regime, as well as identify traces of the interactions between the upper atmosphere and the planet interior. Indeed, the proximity (26.65\,pc) and size (0.182\,R$_\odot$) of its M6-type stellar host \citep{peterson_temperate_2023} makes the transmission spectroscopy metric of the planet the highest amongst sub-400\,K sub-Neptunes.


We used the NIRSpec PRISM mode for our Bright Object Time Series (BOTS \citep{birkmann_near-infrared_2022}) observation to obtain the transmission spectrum of the planet over the full 0.7-5.2 $\mu$m spectral range in one single transit observation. This spectral range covers molecular absorption bands from all the main absorbers in the near-infrared, e.g., H$_2$O, CH$_4$, CO$_2$, NH$_3$, H$_2$S, SO$_2$, CO, allowing us to assess the abundance of each of these species and infer the chemical regime of the atmosphere of LP~791-18\,c. Our time series consisted of 8,175 integrations of 6 groups each over the SUB512 subarray (512 by 32 pixels), yielding a 1.6-s integration time in the NRSRAPID readout pattern. The observation covered 2 hours of pre-transit baseline, the 70-minute transit, and 30 minutes of post-transit baseline. We decided to use 6 groups per integration --- even though it led to saturation on the detector between 1-2 $\mu$m --- in order to optimize the signal-to-noise at longer wavelengths, where molecules such as CH$_4$, CO$_2$ and CO offer their most important opacity. This choice led to an observing efficiency of 71\% compared to 33\% had we performed non-saturated integrations of 2 groups. Our strategy effectively allowed us to collect more than twice as many photons, thus increasing our transit depth precision over most of the wavelength range by a factor 1.47. 

\textbf{Analysis}~~

\begin{figure}[t!]
\begin{center}
\includegraphics[width=0.99\linewidth]{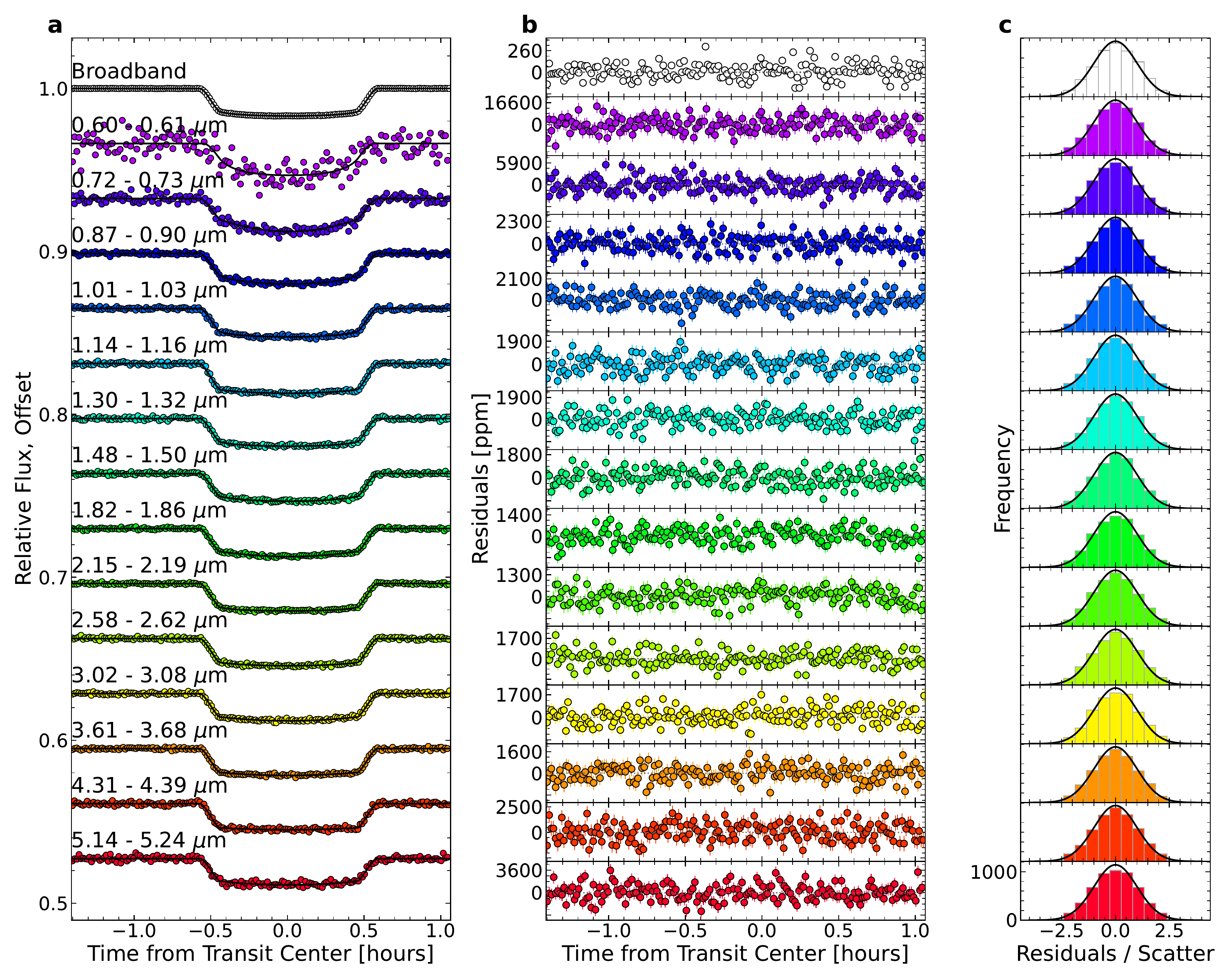}
\end{center}
\caption{\textbf{JWST NIRSpec PRISM broadband and spectrophotometric light-curve fits of the LP~791-18\,c transit observation.} \textbf{a.} Normalized and systematics-corrected transit observations (points) for the broadband light curve (white) and for a sample of 14 spectroscopic bins (colored), along with their respective best-fit models (dark lines). The multiple light curves are plotted with an offset of 0.03 in relative flux. Each light curve is labelled with the edges of the corresponding spectroscopic bin. The observations are displayed in bins of 80\,s for visual purposes, but the full unbinned time array is used for all analyses. \textbf{b.} Residuals from the best-fit model light curves. From the vertical axis of the panels, we observe a variation of the random photometric scatter in the light curves, which is due to the wavelength-dependent stellar photon flux and instrument throughput. \textbf{c.} Histograms of the non-binned residuals divided by the photometric scatter. The residuals follow the expected Gaussian distributions (black curves). }
\label{fig:rainbow}
\end{figure}

We reduced the uncalibrated NIRSpec PRISM data using a custom reduction algorithm (see Methods) built around the STScI pipeline. Compared to previous works \citep{ahrer_identification_2022, rustamkulov_early_2023}, we performed further detector-level corrections such as 1/$f$ correction at the group level and nonuniform-illumination-nonlinearity (NIN) correction (Benneke et al., under review, see Methods). The time series of stellar spectra was then extracted using Stage 3 of the \texttt{Eureka!} framework \citep{bell_eureka_2022}, and the light-curve fitting was performed using the ExoTEP \citep{benneke_spitzer_2017, benneke_sub-neptune_2019, benneke_water_2019} framework. For saturated pixels, we only use the groups that are not saturated for the ramp fitting step. Despite this, the light curves extracted in this region show enhanced correlated noise, and yield transit depths that deviate from the rest of the spectrum.  This is in line with the recent benchmark study of WASP-39\,b which showed that the saturated part of the NIRSpec/PRISM transit spectrum was not in agreement with the overlapping NIRISS SOSS spectrum \citep{carter_benchmark_2024}. Saturation involves a complex interplay of non-linearity, few-groups ramp fitting and pixel bleeding, possibly leading to biased inferences when considering these measurements in atmospheric analyses. We thus conservatively reject all light curves in the 1.0-1.8\,$\mu$m region from our analyses. Outside the saturated region, the broadband and spectroscopic light curves are mostly free of instrumental systematics, and we only use a linear slope over the time series as our instrumental systematics model (Fig. \ref{fig:rainbow}). We further produce an independent reduction of the data using the ExoTEDRF framework and find that it produces a consistent transmission spectrum (Extended Data Fig. \ref{fig:NIN}).


Because M dwarfs such as LP 791-18 are known to be active and exhibit intense stellar flares that can affect JWST time series \citep{howard_characterizing_2023}, we examined the H$\alpha$ emission of the star and found that it remained quiet throughout the observation (Extended Data Fig. \ref{fig:diagnostics}). However, the light curves display a ``bump'' right after the centre of the transit which we attribute to a spot-crossing event due to the decreasing amplitude of the feature with wavelength. We model this signal using two independent methods: a Gaussian spot-crossing model, and the \texttt{spotrod} \citep{beky_spotrod_2014} tool (see Methods). Both analyses result in consistent transmission spectra, and light-curve residuals free of correlated noise (Extended Data Fig. \ref{fig:spot-crossing}). Finally, we move forward with the full pixel resolution transmission spectrum from the analysis with the Gaussian spot-crossing model since it has fewer correlations between parameters than \texttt{spotrod} (see Methods). We opt to use the full-resolution spectrum for our analyses in order to preserve all the available wavelength information and to avoid aliasing when binning the spectrum.

\textbf{Results}~~

\begin{figure}
\centering
   \includegraphics[width=0.8\linewidth]{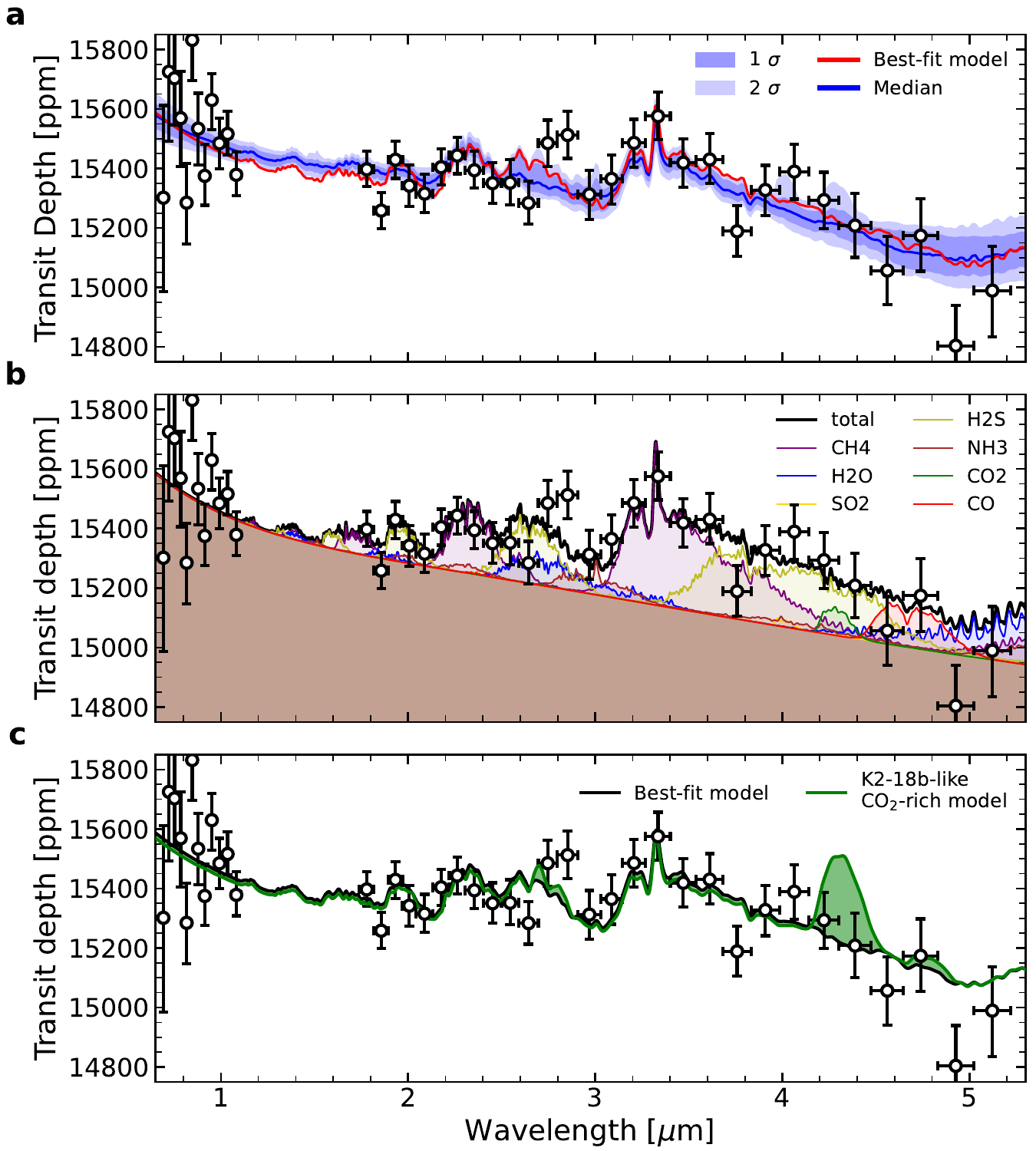}
\caption{\textbf{JWST NIRSpec/PRISM transmission spectrum of the temperate sub-Neptune LP~791-18\,c.} \textbf{a.} The transit spectrum of LP~791-18\,c (black points with 1$\sigma$ error bars) binned at a resolution of R=25 is shown with our model transmission spectra constraints from the nested sampling free chemistry atmosphere retrieval (blue). The dark blue and light blue shaded regions show the 1$\sigma$ and 2$\sigma$ Bayesian credible intervals from the atmosphere retrieval. The best-fitting model is shown in red and the median of our samples is shown in blue. \textbf{b.} Molecular contributions to the retrieved best-fit transit spectrum of LP~791-18\,c. Contributions of CH$_4$, H$_2$O, H$_2$S, SO$_2$, NH$_3$, CO$_2$ and CO are shown in purple, blue, chartreuse, yellow, brown, green and red. The brown region shows the opacity of the aerosols. \textbf{c.} The best-fitting transit spectrum of LP~791-18\,c (black) is compared with a K2-18\,b-like atmosphere model (green). The CO$_2$-rich model is produced from the best-fit model, to which we add CO$_2$ in order to reach the same CO$_2$/CH$_4$ ratio as for K2-18\,b \citep{madhusudhan_carbon-bearing_2023}. The transit spectrum is characterized by hazes that are opaque at short wavelengths with a fading opacity past 2 $\mu$m, by the methane absorption features at 2.3 and 3.3 $\mu$m, and by the absence of the 4.4 $\mu$m CO$_2$ absorption band.}
\label{fig:spectrum}
\end{figure}

\begin{figure}
\centering
   \includegraphics[width=0.24\linewidth]{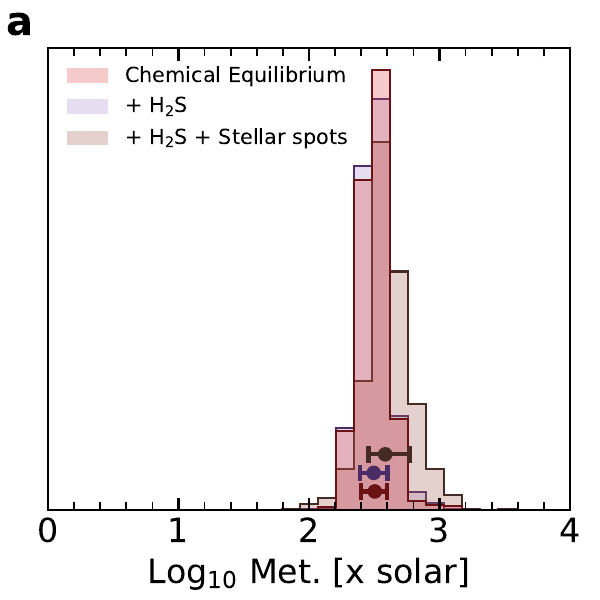}
   \includegraphics[width=0.24\linewidth]{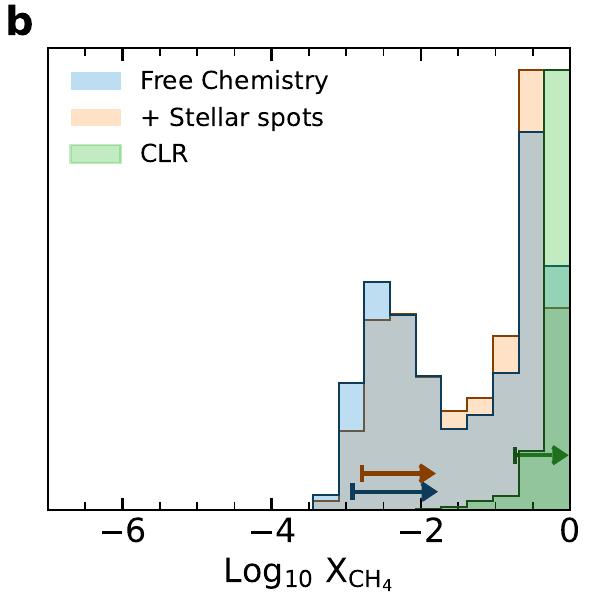}
   \includegraphics[width=0.24\linewidth]{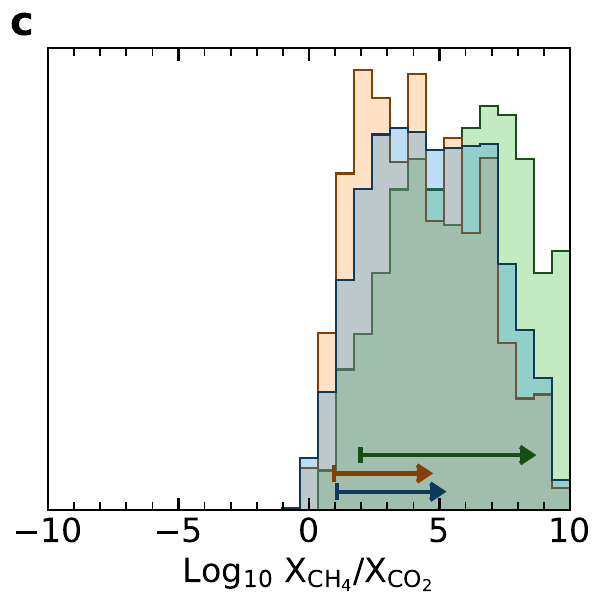}
   \includegraphics[width=0.24\linewidth]{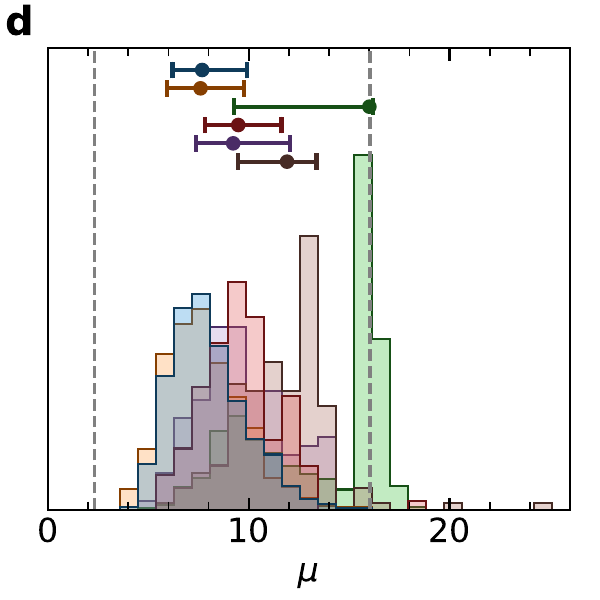}
   
\caption{\textbf{Measured atmospheric properties of LP~791-18\,c.} \textbf{a.} Retrieved posterior probability distributions for the metallicity of LP~791-18\,c's atmosphere based on the suite of chemically consistent retrievals (standard in red, with H$_2$S in purple, with H$_2$S and stellar spots in brown). \textbf{b-c.} Retrieved posterior probability distributions for the abundance of CH$_4$ and the CH$_4$-to-CO$_2$ abundance ratio of LP~791-18\,c's atmosphere based on the suite of free retrievals (standard in blue, with stellar spots in orange, and with the centred-log-ratio parameterization in green). \textbf{d.} Retrieved posterior probability distributions for the mean molecular weight $\mu$ of the atmosphere based on all atmosphere retrievals. The mean molecular weights of H$_2$/He (2.3 amu) and pure CH$_4$ (16 amu) atmospheres are shown as grey dashed lines. In all panels, the 1$\sigma$ Bayesian credible regions or 2$\sigma$ lower limits are shown as bold data points and arrows. While some differences exist between the measurements obtained from the different retrieval parameterizations and priors, the suite of retrievals performed on the spectrum of LP~791-18\,c all depict methane-rich atmospheres.}
\label{fig:measurements}
\end{figure}

The NIRSpec PRISM transmission spectrum of LP~791-18\,c (Fig. \ref{fig:spectrum}) reveals statistically significant (4.47$\sigma$) methane absorption in the 2.3 and 3.3 $\mu$m vibrational bands. These absorption signatures are accompanied by the presence of a strong continuum opacity due to hazes (5.36$\sigma$) that become increasingly transparent towards 5 $\mu$m. Such continuum opacity was not observed for previously studied temperate sub-Neptunes \citep{madhusudhan_carbon-bearing_2023, benneke_jwst_2024, holmberg_possible_2024}. In addition, despite sufficient sensitivity, the data show no molecular absorption of CO$_2$ at 4.3 $\mu$m, providing an overall ``slanted" shape to the transmission spectrum. This particular continuum opacity profile is best matched by a combination of small-particle hazes and Mie scattering clouds \citep{benneke_sub-neptune_2019} in the atmosphere of LP~791-18\,c.
Quantitatively, atmospheric retrievals considering free chemistry along with haze and Mie cloud opacities constrain the methane abundance in LP~791-18\,c's atmosphere to be above the 0.1\% level ($\log X_{\mathrm{CH_4}}>-2.89$ at 2$\sigma$, Fig. \ref{fig:measurements}). Our transmission spectrum does not show CO$_2$ absorption and constrain its abundance to below the percent level ($\log X_{\mathrm{CO_2}}<-1.97$ at 2$\sigma$, see Extended Data Table \ref{tab:mol_constr}). 


We inspect the overall metal enrichment of the atmosphere of LP~791-18\,c using chemically consistent retrievals. Here, the molecular composition of the atmosphere is parameterized via the metallicity and the carbon-to-oxygen (C/O) ratio (Extended Data Fig. \ref{tab:mol_constr}). As for the free-chemistry retrievals, hazes and Mie clouds are simultaneously considered, and the conclusions are unchanged by the addition of an out-of-equilibrium species like H$_2$S (see Methods). We find the atmosphere of LP~791-18\,c to be highly enriched in metals, with a metallicity in the 250--400$\times$ solar range (Fig. \ref{fig:measurements}). Although no oxygen bearing species is directly detected in the spectrum of LP~791-18\,c, we obtain two-sided constraints on the metallicity in the atmosphere. This is because, apart from scenarios with much more oxygen than carbon (leading to atmospheres without methane, C/O$>$0.16 at 2$\sigma$), methane remains the dominating opacity source over much of the metallicity-temperature parameter space, until CO$_2$ becomes abundant at high metallicities. We also expect H$_2$O to be present at chemical equilibrium in the atmosphere of LP~791-18\,c for all metallicities consistent with the data; however, the water is not detected since methane is much more opaque than water over the observed wavelength range, given the low temperature of the planet \citep{benneke_jwst_2024}. In addition, the hazes are hiding most molecular bands shortwards of 2\,$\mu$m. While unocculted stellar spots could lead to contamination of transmission spectra of exoplanets via the transit light-source (TLS \citep{rackham_transit_2018, rackham_transit_2019}) effect, we find that our results are independent of and robust to the inclusion of stellar contamination in the retrievals (see Methods and Extended Data Fig. \ref{fig:ChemEquiCorner}).

\textbf{Discussion}~~

The 0.7--5.2 $\mu$m transmission spectrum of LP\,791-18\,c presented in this work shows clear differences to the other two low-temperature sub-Neptunes characterized with JWST to date: K2-18\,b \citep{madhusudhan_carbon-bearing_2023, schmidt_comprehensive_2025} and TOI-270\,d \citep{benneke_jwst_2024, holmberg_possible_2024}. That is despite LP\,791-18\,c being intermediate in equilibrium temperature and planet radius. The transit spectra of K2-18\,b and TOI-270\,d display mostly clear atmospheres, with multiple bands of CH$_4$ detected, and with CO$_2$ absorption detected at 4.3\,$\mu$m (although the detection of CO$_2$ in K2-18\,b's atmosphere is debated \citep{madhusudhan_carbon-bearing_2023, schmidt_comprehensive_2025}). Since LP\,791-18\,c has an equilibrium temperature between that of K2-18\,b and TOI-270\,d, its distinct transmission spectrum points to an intrinsic diversity in the sub-Neptune population, both in terms of chemistry and aerosol regimes. In the following, we explore possible causes for these differences, and explanations for the nature of LP~791-18\,c. We start by considering and contrasting the chemical regime of the planet, and then examine its aerosols versus the rest of the sub-Neptune population. 

On one hand, the detection of methane along with the non-detection of CO$_2$ in the atmosphere of LP~791-18\,c is consistent with a metal-enriched atmosphere close to chemical equilibrium. Chemical equilibrium predicts that methane and water should be the main absorbers for atmospheres in the temperate regime, and that CO$_2$ should not be abundant. This is in line with our analysis of the transit spectrum of LP~791-18\,c. We find methane to be the dominant molecular absorber, and we do not detect CO$_2$. As explained above, we do not find constraints on the water abundance since it is hidden under the methane and haze opacities. Furthermore, we do not detect direct signs of disequilibrium chemistry, either in the form of quenched or photochemically produced species. Hence, a metal-rich (250--400 $\times$ solar), deep atmosphere near chemical equilibrium is the scenario that explains the data at hand in the most natural way.

On the other hand, the recent detections of CO$_2$ in the transmission spectra of TOI-270\,d and K2-18\,b were proposed to be the result of vertical mixing in highly metal-enriched atmospheres \citep{benneke_jwst_2024}. This led to the idea that all sub-Neptunes could have miscible envelopes, where the upper atmosphere is fully miscible with the supercritical water in the planet interior. This allows the CO$_2$, produced in the deeper and hotter regions of the envelope, to be quenched into the photosphere. In order to explore this scenario for LP~791-18\,c, we fully inspect the effect of disequilibrium chemistry by generating a grid of SCARLET radiative-convective atmosphere models of LP~791-18\,c, for which the vertical mixing is modelled using the VULCAN framework \citep{tsai_vulcan_2017} (with K$_{zz}=10^4$ cm$^2$/s \citep{benneke_jwst_2024}). The models in the grid are produced for varying Bond Albedo and atmospheric metallicities, and allow us to track the evolution of the CH$_4$-to-CO$_2$ abundance ratio (CH$_4$/CO$_2$) with these parameters. Our analysis indicates that only at metallicities above 500 $\times$ solar would we expect CO$_2$ to become similarly or more abundant than CH$_4$, and this value rises with Bond albedo (rises with decreasing temperatures), which could be non-zero on LP~791-18\,c given the hazes in its atmosphere (Extended Data Fig. \ref{fig:miscibleEnv}). 

Our grid exploration reveals that LP~791-18\,c is also potentially consistent with a miscible envelope sub-Neptune, similar to TOI-270\,d. The metallicity of 200--450$\times$solar measured in the chemically consistent retrieval is within the methane dominated regime of the grid of models (Extended Data Fig. \ref{fig:miscibleEnv}). This is also seen in the CH$_4$/CO$_2$ measured for LP~791-18\,c (log CH$_4$/CO$_2  > 1.07 \ (2\sigma)$) and for TOI-270\,d (log CH$_4$/CO$_2  = 0.95^{+0.69}_{-0.52}$ \citep{benneke_jwst_2024}), which are marginally consistent. In that scenario, the lower temperature of LP~791-18\,c compared to TOI-270\,d could lead to a smaller CO$_2$ abundance (larger CH$_4$/CO$_2$ ratio) in the photosphere, and the hazes in the atmosphere of LP~791-18\,c could hinder its detection. The story, however, becomes more complicated when one considers that carbon dioxide was measured to be more abundant than methane on the cooler K2-18\,b (log CH$_4$/CO$_2$  = -0.3 $\pm$ 0.8)\citep{madhusudhan_carbon-bearing_2023}. Assuming a similar composition, this is the opposite of what is expected for a cooler planet, since cooler temperatures should impede CO$_2$ production. It thus follows that a miscible envelope scenario along with temperature transitions is not sufficient to uniquely describe the observed chemical inventories of all three planets. One should note that, while this detection of CO$_2$ in the atmosphere of K2-18\,b has recently been questioned by a reanalysis of the single NIRISS SOSS and NIRSpec G395H transits with multiple reduction methods \citep{schmidt_comprehensive_2025}, the analysis of 4 additional NIRSpec transits of K2-18\,b (GO 2372) confirms the CO$_2$ detection (Hu et al., in prep.). 

Different formation histories might explain the differences in atmosphere composition of TOI-270\,d, LP~791-18\,c, and K2-18\,b. If, despite their varying upper-atmosphere CO$_2$ abundances, the three planets are all miscible envelope sub-Neptunes, then it could be that TOI-270\,d and K2-18\,b have more oxidative atmospheres, with higher oxygen contents (higher O/H and lower C/O), favouring CO$_2$ production \citep{yang_chemical_2024}. This could be explained by formation histories in more or less wet (H$_2$O rich) or dry (less H$_2$O, more H$_2$) environments. Another possibility is that the planets are fundamentally different, with highly different amounts of rock and ice in the interior. For instance, the CO$_2$ observed on K2-18\,b has been proposed to be the consequence of a thin atmosphere atop a liquid water surface \citep{madhusudhan_carbon-bearing_2023} or a magma surface \citep{shorttle_distinguishing_2024} which both can act as a source for CO$_2$ production. For LP~791-18,c, the CO$_2$-to-CH$_4$ ratio (CO$_2$/CH$_4$) observed in the atmosphere can similarly be used to infer the bulk composition and formation history. The measurement of an upper atmosphere enriched in CH$_4$ compared to CO$_2$ could then indicate a bulk composition depleted in water in comparison to H$_2$ \citep{yang_chemical_2024}. It was recently shown that for low-temperature sub-Neptunes, the upper atmosphere CO$_2$/CH$_4$ is correlated with the bulk H$_2$O-to-H$_2$ ratio (H$_2$O/H$_2$), or the water enrichment in the deep interior \citep{yang_chemical_2024}. For an upper atmosphere with CH$_4$ as the main carbon-bearing species, we thus expect the bulk composition of the planet to be richer in H$_2$ than in H$_2$O. This in turn hints at a formation in a dry environment, inside the water ice-line, where the planet would have accreted mostly H$_2$ compared to H$_2$O ice (low O/H ratio) providing an explanation for the CH$_4$-rich and CO$_2$-free atmosphere of LP~791-18\,c. 

\begin{figure}
\centering

\sbox{\bigpicturebox}{%
  \begin{subfigure}[b]{.50\textwidth}
  \scalebox{1}[1.2]{\includegraphics[width=\textwidth]{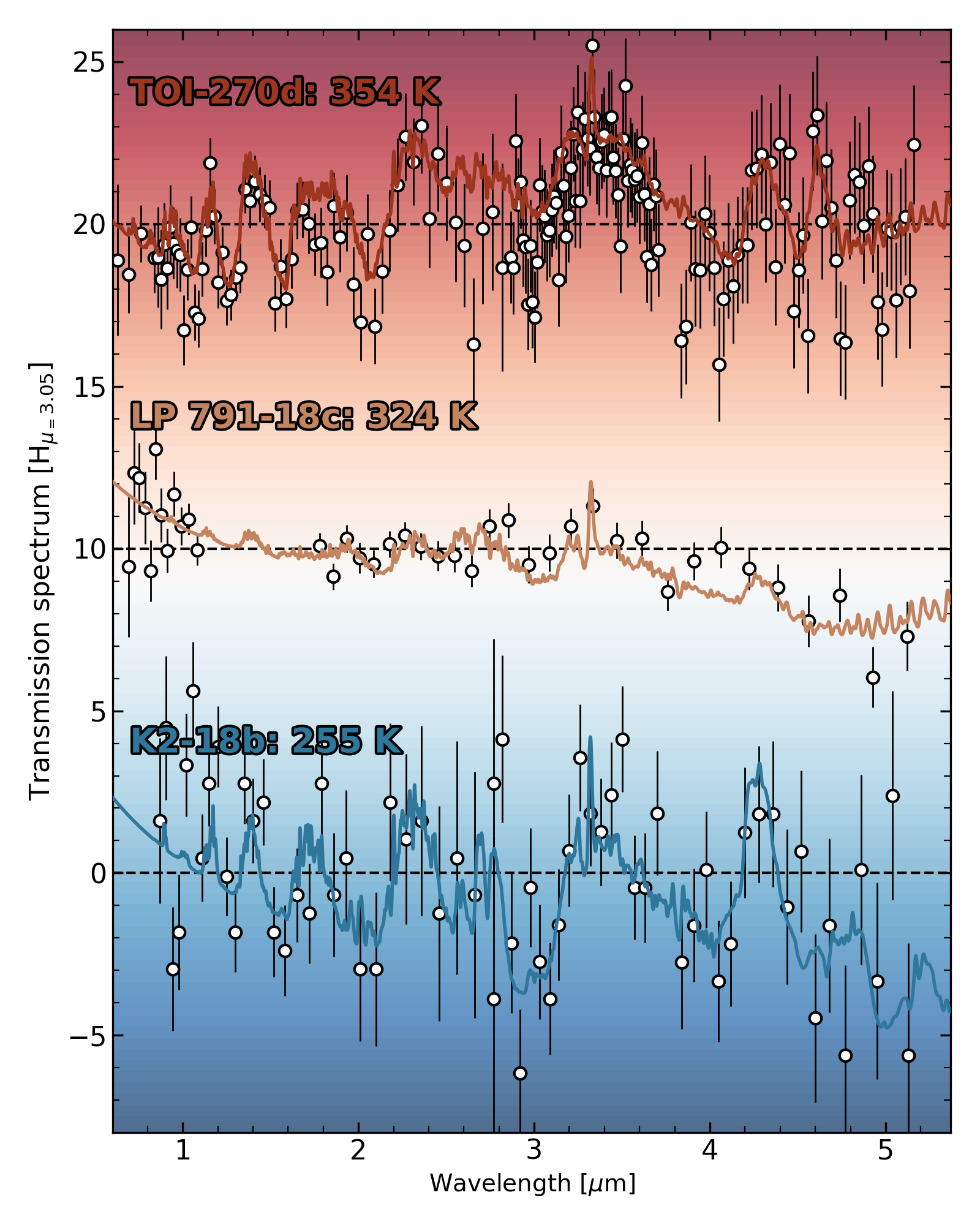}}%
\end{subfigure}
}

\usebox{\bigpicturebox}\hfill
\begin{minipage}[b][\ht\bigpicturebox][s]{.45\textwidth}
\begin{subfigure}{\textwidth}
\includegraphics[width=\textwidth]{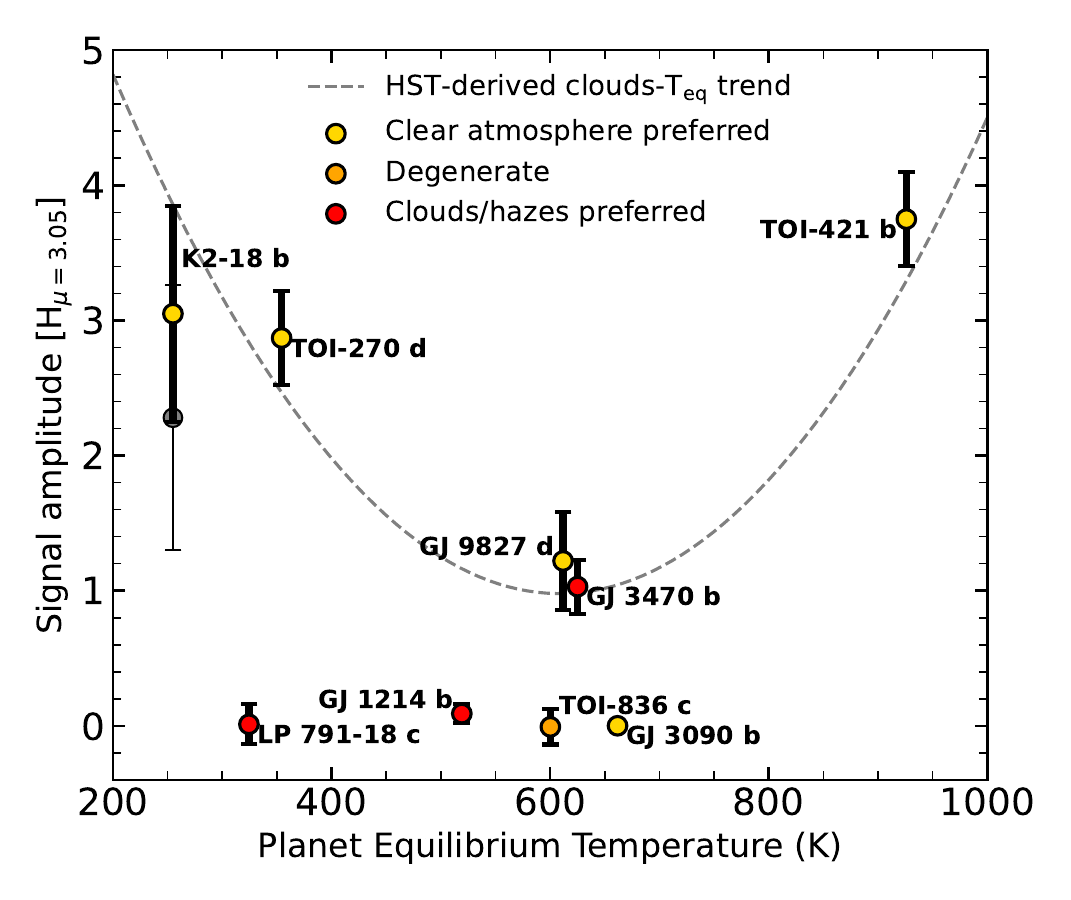}
\end{subfigure}\vfill
\begin{subfigure}{\textwidth}
\includegraphics[width=\textwidth]{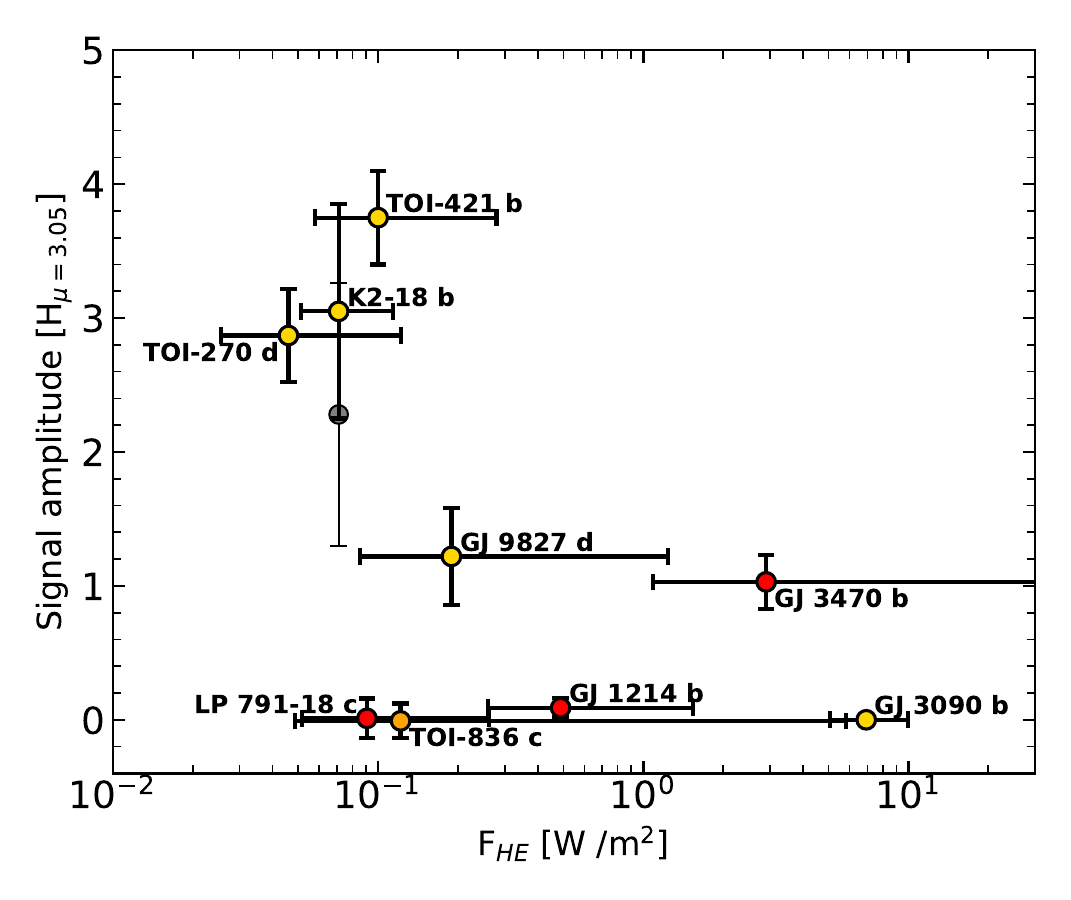}
\end{subfigure}


\end{minipage}



   
\caption{\textbf{Diversity in sub-Neptune transmission spectra.} \textbf{a.} The JWST transmission spectra of TOI-270\,d \citep{benneke_jwst_2024}, K2-18\,b \citep{madhusudhan_carbon-bearing_2023}, and LP~791-18\,c are shown with their 1$\sigma$ error bars and respective best-fit models. The transmission spectra are shown in units of scale heights for a 3.05 amu mean molecular weight, and are color coded as a function of their equilibrium temperature (for A$_B$=0.3). \textbf{b.} Retrieved atmospheric signal strength with 1$\sigma$ error bars for JWST-characterized sub-Neptunes as a function of their equilibrium temperature (for full heat redistribution and 0.3 Bond albedo). Sub-Neptunes are colored depending on whether the atmosphere characterization revealed the upper atmosphere to be mostly clear (yellow), mostly cloudy or hazy (red), or whether it is still unknown (orange). The grey point shows the retrieved signal strength of K2-18\,b when no detector offsets are allowed in the retrieval. The equilibrium temperature trend proposed to be a consequence of clouds from the HST survey of sub-Neptunes is shown as the dashed line. \textbf{c.} The retrieved atmospheric signal strength for characterized sub-Neptunes as a function of the high-energy (X-ray to UV) irradiation of each planet are shown with their respective 1$\sigma$ error bars. }
\label{fig:population}
\end{figure}

Outside of the chemical composition, the main difference between the transmission spectrum of LP~791-18\,c and that of the other temperate sub-Neptunes studied to date is the prominence of hazes. While preference for some amount of hazes and clouds was found in the analyses of the transmission spectra of K2-18\,b and TOI-270\,d, their effects on the opacity profile remain marginal, allowing the detection of all CH$_4$ bands across the 0.6-5.0\,$\mu$m range \citep{madhusudhan_carbon-bearing_2023, benneke_jwst_2024}. Conversely, hazes represent the dominating source of opacity in the transmission spectrum of LP~791-18\,c, greatly muting the spectral features (Fig. \ref{fig:population}). Quantitatively, the Rayleigh enhancement factor due to hazes measured on LP~791-18\,c ($\log c_{\mathrm{Haze}} = 5.16^{+0.5}_{-0.7}$) is two and three orders of magnitude larger than that observed on K2-18\,b ($\log c_{\mathrm{Haze}} = 2.7^{+0.8}_{-3.8}$) and TOI-270\,d ($\log c_{\mathrm{Haze}} = 1.9^{+0.5}_{-6.2}$), respectively (see Methods).  Recent trends based on a HST survey of sub-Neptune atmospheres had indicated that low-temperature sub-Neptunes appeared mostly free of aerosols in their upper atmospheres compared to their counterparts in the 500--800\,K regime \citep{brande_clouds_2024}. LP~791-18\,c constitutes a strong outlier to that trend, by providing a signal amplitude consistent with zero at 1.4 $\mu$m despite its low equilibrium temperature (Fig. \ref{fig:population}). We further compare the HST-derived trend in sub-Neptune absorption signals to the updated population of JWST-characterized sub-Neptunes (Fig. \ref{fig:population}). Out of the nine sub-Neptunes observed in transmission with JWST as of now, we find that four are inconsistent with the clouds-T$_{eq}$ trend of ref.\citep{brande_clouds_2024}, i.e., sub-Neptunes with muted transmission spectra are found across the temperature range. Moreover, while GJ 9827\,d and GJ 3090\,b appear consistent with the clouds-T$_{eq}$ trend, with low-amplitude absorption bands in the 600--700\,K range; the reason for their muted transit spectra is their high mean-molecular-weight envelopes rather than the presence of aerosols \citep{piaulet-ghorayeb_jwstniriss_2024} (Ahrer et al., under review). Our study of LP~791-18\,c, along with the first wave of JWST sub-Neptune characterization, highlights the diversity in the spectra of sub-Neptune exoplanets.


The host star spectral type of LP~791-18\,c (M6 compared to M3 for K2-18\,b and TOI-270\,d) could provide a potential explanation for the distinct aerosol regime observed on the planet. Varying levels of high-energy irradiation can lead to different photochemical reactions in sub-Neptune upper atmospheres, which in turn can result in different rates of formation of hazes. We investigate the sub-Neptune atmosphere signal strengths as a function of incoming high-energy flux (X-ray to UV) calculated from the stellar masses and ages \citep{owen_evaporation_2017, jackson_coronal_2012}. We find that, along this high-energy flux axis, the three low-temperature sub-Neptunes are ordered consistently with the amount of hazes in their atmospheres (Fig. \ref{fig:population}). While there is no clear trend yet, especially considering the large uncertainties on the high-energy fluxes (stemming from the uncertain stellar ages), none of the sub-Neptunes with large features in their transmission spectra are found in the highly-irradiated regime (Fig. \ref{fig:population}). In the future, expanding the sample of characterized (and mainly, JWST-characterized) planets has the potential to reveal key correlations between the irradiation environment of sub-Neptunes and the aerosols photochemically-produced in their upper-atmosphere.

Combining the presence of hazes with the metal-enriched envelope discussed earlier depicts a somewhat self-consistent picture for the atmosphere of LP~791-18\,c. High photochemical reaction rates (that stem from high-energy irradiation) and high atmospheric metallicities, are both properties that favour the efficient formation of haze precursors \citep{morley_thermal_2015, gao_hazy_2023}. In the context of a CH$_4$-rich atmosphere, CH$_4$-derived tholin hazes, such as those found on Titan \citep{waite_process_2007, robinson_titan_2014}, could naturally explain the hazy nature of the planet, especially as they are expected to have a decreasing optical depth with wavelength, similar to what is observed in the spectrum of LP~791-18\,c \citep{morley_thermal_2015, gao_aerosol_2020}. Sulfur hazes are another explanation for the nature of the aerosols present in the atmosphere of LP~791-18\,c, and could stem from the photolysis of H$_2$S to form S$_8$, or H$_2$SO$_4$ \citep{hu_photochemistry_2013}. We explored the absorption cross-sections of known haze precursors  (HCN, C$_2$H$_2$, C$_2$H$_4$, C$_2$H$_6$, SO$_2$, and SO, see Methods)\citep{waite_process_2007,miller-ricci_kempton_atmospheric_2012, moses_compositional_2013, hu_photochemistry_2013}, but did not find evidence for their presence in the atmosphere of LP~791-18\,c. Further observations of LP~791-18\,c are needed to characterize the type of hazes in its atmosphere. In particular, transit observations with the MIRI LRS mode could extend the transmission spectrum of LP~791-18\,c up to 12\,$\mu$m, potentially providing a better view of the fading cloud opacity and potential cloud features \citep{grant_jwst-tst_2023}. Moreover, by probing a wavelength range where the cloud opacities are expected to fade, such observations have the potential to provide a clearer view at the molecular absorption bands (such as CH$_4$, H$_2$O, SO$_2$) and refine the chemical constraints on the atmosphere of LP~791-18\,c.

\newpage
\begin{methods}

\subsection{Observations} 
We observed a single transit of the temperate sub-Neptune LP~791-18\,c with the James Webb Space Telescope (JWST) using the Near-Infrared Spectrograph (NIRSpec) in PRISM mode. We used NIRSpec's Bright Object Time Series (BOTS) mode with the NRSRAPID readout pattern, the SUB512 subarray and the S1600A1 slit. The exposure consisted of 8175 integrations of 6 groups each (for an integration time of 1.6\,s), and covered the full planetary transit along with 2.5\,h of baseline. We achieved an observing efficiency of 71\% by wilfully saturating the detector in the 1-2 $\mu$m region (see below). 

\subsection{\textit{JWST}/NIRSpec PRISM data reduction summary} 
We reduced the JWST/NIRSpec PRISM observations using a combination of a custom Stage 1 group-level routine and of the \texttt{Eureka!}\citep{bell_eureka_2022} framework (Extended Data Fig. \ref{fig:reduction}). Our Stage 1 is built around the STScI pipeline and processes the data up to ramp-fitting. Our custom version adds numerous corrections to the standard reduction pipeline, such as group-level 1/$f$ background subtraction, and nonuniform-illumination-nonlinearity (NIN) correction. The data calibration and the extraction of stellar spectra, Stages 2 and 3, are then performed with \texttt{Eureka!}.

We start our data analysis from the raw uncalibrated data and follow the standard \texttt{jwst} pipeline with the following modifications. We skip the jump detection step as it greatly overestimates the number of outliers in the frames, which leads to well-behaved pixels being flagged as outliers. This was also observed for other NIRSpec PRISM time series using a similarly small number of groups \citep{rustamkulov_early_2023}. Outliers are rather flagged in the spectral extraction step of our data reduction. We also add a group-level background subtraction step in our Stage 1 routine. Performing this step before ramp-fitting is crucial to remove the 1/$f$ noise \citep{rustamkulov_early_2023}, as removing the background after ramp fitting leads to larger scatter. In that step, we compute and subtract a mean to each detector column (using the top and bottom 6 pixels of each column to evaluate the background), and repeat this for each group of each integration. Pixels that are 3$\sigma$ outliers in the column-by-column fits are ignored.

\subsection{Non-linearity treatment} 
We use the recently documented NIN correction (Benneke et al., under review) to correct for the non-linear response in each pixel. The NIN effect is a consequence of the discrepancy between the uniform illumination of the detector that led to the STScI-provided non-linearity coefficients, and the highly non-uniform illumination of the detector in PRISM mode. This effect, observable in the unexpected and substantial variations of the pixel-by-pixel transit depth along the cross-dispersion direction, leads to systematic over- and under-corrections of the non-linearity depending on the pixel row (Benneke et al., under review). 

Our NIN-correction thus works by first performing the standard \texttt{jwst} non-linearity step. This standard non-linearity correction leaves systematic residuals between the best-fitting ramp and the corrected reads up-the-ramp, and we fit these residuals with a quadratic polynomial, obtaining a residual quadratic term for each pixel, which we observe to vary systematically and smoothly as a function of illumination. We thus fit a second order polynomial to these residual quadratic terms as a function of the median slope in the central row, and use these fitted coefficients to correct the non-linearity for all groups and all integrations. By fitting a quadratic curve to the residual quadratic terms, we effectively smooth-out our NIN correction by studying the residuals-to-illumination behaviour of the detector instead of correcting pixel-by-pixel. In this step, we do not use pixels whose illumination approaches saturation. 

The NIN correction performed here corrects a broad trend in the transit spectrum of LP~791-18\,c, which is most important around 2\,$\mu$m (Extended Data Fig. \ref{fig:NIN}).  The correction is very similar to that of TRAPPIST-1\,g's transit observation, which was performed under the exact same observational settings (NIRSpec/PRISM, 6 groups per integration including partial saturation in the 1-2 $\mu$m region; Benneke et al., under review). As such, we are confident that the NIN correction yields the best known non-linearity correction for PRISM transit observations and use it for our LP~791-18\,c transmission spectroscopy analyses.

\subsection{Saturation and ramp fitting}
We use six groups per integration for our time series observation in order to increase the observing efficiency and subsequently the signal-to-noise over most of the detector, at the cost of saturation in the brightest pixels (Extended Data Fig. \ref{fig:reduction}). In order to recover transit depth measurements from the saturated region of the spectral trace, we ignore saturated groups, as flagged by the standard STScI routine, when performing the ramp fitting. Despite this, we find that the affected pixels produce extracted light curves which have important correlated noise, and which yield discrepant transit depths when compared to the rest of the wavelength range. While more complex methods could be used to try and recover the signal in those pixels, it was shown using the WASP-39\,b benchmark data set that PRISM saturated regions, regardless of the methods used, are unable to reproduce the same signal as overlapping, unsaturated instruments \citep{carter_benchmark_2024}. It thus makes it risky to use the corresponding light curves and derived transit depths for atmosphere analyses. For this reason, we decide to reject this saturated region (1-1.8\,$\mu$m) from our analyses.  

\subsection{Data calibration}
The integration-level data calibration is performed with Stage 2 of the \texttt{Eureka!} pipeline, which is effectively a wrapper around the standard STScI-provided Stage 2. We use all default settings but skip the photometric calibration and flat-fielding steps, as absolute flux units are not needed for transit depth measurements. An example integration-level calibrated frame is shown in Extended Data Fig. \ref{fig:reduction}.

\subsection{Extraction of stellar spectra}
We use Stage 3 of the \texttt{Eureka!} framework to perform the extraction of the time series of stellar spectra. For this step, we use all 32 pixels of the subarray in the cross-dispersion direction, but restrict the aperture to pixels 14-495 in the dispersion direction, as pixels outside this range receive negligible illumination. We start by performing an integration-level background subtraction using the same method as for the group-level step. Since most of the background signal was already removed at the group level, this step does not have a large effect on the results and is mostly there for redundancy. The centre of the trace is obtained by fitting a Gaussian curve to the sum of the pixels along detector rows, and the background is estimated using all pixels that are at least 8-pixels away from that central row. Outliers are flagged and ignored prior to this step by running two iterations of 10-$\sigma$ clipping in time, as well as a single iteration of spatial 3-$\sigma$ clipping. We then perform a median-weighted optimal extraction on an aperture restricted to the 7 pixel rows above and below the centre of the trace (Extended Data Fig. \ref{fig:reduction}). The median-weighted optimal extraction profile is also corrected for consistently bad pixels (for instance, hot pixels), using the median of the data quality arrays and giving zero weight to bad pixels, which helps in obtaining smooth stellar spectra (Extended Data Fig. \ref{fig:reduction}). For each integration, we record the vertical shift of the trace on the detector using cross-correlation with the first integration, and we record the width of that cross-correlation function (Extended Data Fig. \ref{fig:diagnostics}).

\subsection{Data reduction with the ExoTEDRF framework}
In order to ensure that our data reduction is robust, we produce an independent analysis of our LP~791-18\,c transit using the ExoTEDRF pipeline \citep{radica_awesome_2023, feinstein_early_2023, radica_exotedrf_2024}, which was recently updated and benchmarked for NIRSpec/G395H observations (Ahrer et al., under review). We thus perform the data reduction closely following the steps outlined in Ahrer et al. (under review). We apply the usual Stage 1 reduction steps up to a few exceptions. First, we skip the reference pixel step. We also perform a group-level 1/f background correction step which removes the median of each detector column, considering pixels that are at least 8 pixels away from the centre on each side, and masking bad pixels. We perform the jump step before ramp fitting in our ExoTEDRF Stage 1 reduction, and add a cosmic ray detection step. 

We apply the standard ExoTEDRF Stage 2 steps to calibrate our integration-level frames, which include running the extract2d step, allowing a wavelength solution correction based on the position of the star in the NIRSpec slit. We also repeat the 1/f background subtraction during Stage 2, and perform an outlier correction step. Before moving to the spectral extraction, we perform a detector-level principal component analysis (PCA) and use these PCA components to obtain the time series of the position of the trace on the detector. Finally, the spectral extraction we perform consists in a standard box extraction using a half-width of 4 pixels. 

Despite the many differences that exist between the steps and corrections applied by our Eureka! and ExoTEDRF reductions, both yield transmission spectra which are in strong agreement (Extended Data Fig. \ref{fig:NIN}). The ExoTEDRF spectrum displays slightly increased channel-to-channel scatter at longer wavelengths, but otherwise the spectra are fully consistent. The exact same light-curve fitting method (described below) was used on both reductions. We are thus confident that our results are robust to the data reduction method.

\subsection{Light-curve preparation}
We construct and fit the NIRSpec/PRISM light curves using the \texttt{ExoTEP} \citep{benneke_spitzer_2017, benneke_water_2019, benneke_sub-neptune_2019} framework. We start from \texttt{Eureka!}'s Stage 3 output, a time sequence of stellar spectra, and build spectrophotometric light curves at full pixel resolution (i.e. one light curve for each pixel column). We create a broadband (white) light-curve by adding the flux from all extracted pixel columns together, except for the 1-2 $\mu$m region where there is saturation. This procedure was demonstrated to decrease the white light curve scatter by 15 to 20\% in observations of TRAPPIST-1\,d which are similarly saturated (Piaulet-Ghorayeb et al., under review). While constructing the spectroscopic and broadband light curves, we record the shift of the trace in the dispersion direction at each integration by cross-correlating the extracted stellar spectra in the wavelength direction. We also record the width of the cross-correlation function at each integration. Saturated pixels are not included in this step to avoid biasing the diagnostic measurements. 


In transit observations of exoplanets around the M star TRAPPIST-1, numerous H$\alpha$ flares were observed \citep{lim_atmospheric_2023, howard_characterizing_2023} and greatly hindered our ability to retrieve well-behaved light-curves and reliable transit depth measurements. We thus monitor the H$\alpha$ activity of LP 791-18 during our time series in order to search for potential flares that could affect the light curves. We use the pixel columns from 0.646 to 0.6625 $\mu$m to estimate the system's H$\alpha$ emission, and use pixels within 0.005 $\mu$m on each side to be the continuum. We divide this spectral band by its maximum and subtract a straight line fitted to the continuum on each side of the H$\alpha$ line. We then integrate the flux in the resulting spectral band for each integration, giving us a time series of the H$\alpha$ activity. We find that contrary to TRAPPIST-1, LP 791-18 remains quiet in terms of H$\alpha$ activity over the full duration of our exposure (Extended Data Fig. \ref{fig:diagnostics}).

\subsection{White-light-curve fitting} 
We use the \texttt{ExoTEP} framework to fit the white-light curve of our LP~791-18\,c transit. We first remove the first 70 minutes of the time series, as they display small-amplitude long-timescale variations, and otherwise do not affect the precision of our fit, given the long baseline before transit. We then remove outliers in trace position and trace width using 4-$\sigma$ clipping, and finally remove 3-$\sigma$ outliers in flux. The light-curves are median-normalized before performing the fitting. 

We find that the white-light curve displays a bump (a rapid increase and decrease in flux) shortly after the middle of the planet's transit. As this event does not correlate with any activity in our diagnostic measurements of the trace position or of H$\alpha$ emission, we postulate that it is due to a spot-crossing event, where the planet temporarily occults a dark cooler spot of the star's surface. We thus fit for this feature using a Gaussian shaped functional over the integration time $t$, defined by 
\begin{equation}
    \mathrm{Model}_{\mathrm{SC}}(t) = A_\mathrm{SC} \, \exp {\frac{-(t - t_\mathrm{SC})^2}{2\, \sigma_{\mathrm{SC}}^2 }},
\end{equation}
where $A_\mathrm{SC}$ is the normalized amplitude of the spot-crossing bump, $\sigma_{\mathrm{SC}}$ is the standard deviation of the Gaussian (directly related to the spot-crossing duration), and $t_\mathrm{SC}$ is the central time of the spot-crossing event (see below for a detailed analysis of the spot-crossing event).

We perform the light-curve fitting on the cleaned white-light curve with a Bayesian analysis using the \texttt{emcee}\citep{foreman-mackey_emcee_2013} package to fully explore the multidimensional posterior distribution of our parameters with the Markov Chain Monte Carlo (MCMC) method. Our astrophysical transit model uses the \texttt{Batman}\citep{kreidberg_batman_2015} package, and we fit for the transit depth, the semi-major axis ($a/R_\star$), the impact parameter ($b$), the transit time ($T_0$) and two quadratic limb darkening coefficients ($u_1$,$u_2$). We add to this model a two-parameter linear slope in time (offset and slope), a photometric scatter parameter, and the three-parameter spot-crossing model. This brings us to a total of 11 parameters in our final white-light curve fit. We use 44 walkers in our MCMC analysis (four times the number of parameters) and run the fit for 9000 steps, well past the convergence of the MCMC chains. Our model correctly fits both the transit signal and the spot-crossing event, returning well-behaved normally distributed scatter in the residuals (Fig. \ref{fig:rainbow}). The retrieved orbital parameters as well as their priors are shown in Extended Data Table \ref{tab:bestfitparams}.

\subsection{Spectroscopic light-curve fitting}
Once the white-light-curve fit has converged, we fix the orbital parameters ($T_0$, $b$, $a/R_\star$) as well as the spot-crossing time and width ($t_\mathrm{SC}$, $\sigma_{\mathrm{SC}}$) to the best-fit values found in the white-light-curve analysis. We then independently perform the same MCMC analyses on each spectroscopic bin (at full spectral resolution, i.e., one bin per pixel column), using 24 walkers, or four walkers for each of the 6 remaining fit parameters. The fits ran for 3000 steps and yielded normally distributed residuals (Fig. \ref{fig:rainbow}). The use of the $(u_1, u_2)$ limb darkening coefficients, and the choice to fit the light curves at full resolution for the atmospheric analyses ensures that our transmission spectrum is free of limb-darkening related biases and that the full spectral information is used in the atmosphere analysis. \citep{coulombe_biases_2024}. We further produce versions of the transmission spectrum at resolving powers of $\lambda/\Delta\lambda$ = 25 and 50, which we use for the figures (for clarity) and the comparisons between analysis methods. Finally, we do not use the transit depths in the $<0.68\,\mu$m range for the atmosphere analyses, as we find that they have much larger error bars and slightly lower transit depths, perhaps indicative of the diminishing stellar emission and instrument throughput.


\subsection{Spot-crossing event fitting with spotrod}\label{sec:SCevent}
In order to physically model the spot-crossing in the transit, we perform an independent light-curve fitting analysis using \texttt{spotrod} \citep{beky_spotrod_2014}. We also produce a transmission spectrum using this transit model to verify the self-consistency of our Gaussian model. Following a similar procedure to ref.\citep{fournier-tondreau_near-infrared_2024}, we construct a broadband light curve by summing the flux from wavelengths 0.65-1.5\,$\mu$m, except that we discard the ill-behaved 1.0-1.2\,$\mu$m regions. We fix the orbital period to 4.9899093\,d \citep{peterson_temperate_2023}, the eccentricity to 0.00008 \citep{peterson_temperate_2023} and the argument of periastron to 0$^{\degree}$. We fit with wide, uninformative priors the transit depth, $T_0$, $b$, $a/R_\star$, $u_1$, $u_2$, the two-parameter linear slope in time and a scalar jitter term. We fit one spot-crossing event, which requires four additional parameters, that is, the spot $x$- and $y$-position ($x_{\rm spot}$, $y_{\rm spot}$), the spot radius $R_\mathrm{spot}$ and the spot-to-stellar flux contrast $F_\mathrm{spot}/F_*$. The spot's position and radius are in stellar radii units with the star's centre at (0, 0). We employ the \texttt{Juliet} light curve fitting tool \citep{espinoza_juliet_2019} with  \texttt{spotrod} for the transit model and \texttt{dynesty} \citep{speagle_dynesty_2020} to sample the parameter space with 2000 live points. We also tested a model with a spot and a facula (bright region) and a model with two spots, but found that the broadband light curve is best-fit by a transit model with one spot crossing-event ($>$\,5\,$\sigma$ preference). Multiple fits of the broadband light curve were performed to establish a most likely solution because the $y$-position of the spot is not well constrained, similarly to ref. \citep{fournier-tondreau_near-infrared_2024}. The most likely best-fit model corresponds to a spot $y$-position of y$_{\rm spot}$ = 0.19$^{+0.14}_{-0.7}$\,R$_*$ as a much lower or higher spot requires a larger spot size that is not supported by current knowledge on the coverage of spots in M dwarf stars \citep[e.g.,][]{rackham_transit_2019}. Following ref. \citep{fournier-tondreau_near-infrared_2024}, we select the set of parameter values with the highest likelihood instead of relying on the medians of the posterior distributions of the broadband light curve fit. This transit model corresponds to a best-fitting spot $x$-position of x$_{\rm spot}$ = 0.56\,R$_*$, $y$-position of y$_{\rm spot}$ = 0.24\,R$_*$ and a radius of the spot R$_{\rm spot}$ = 0.19\,R$_*$. This transit model is overplotted in the right panel of the Extended Data Fig. \ref{fig:spot-crossing}, and we obtained a reduced Chi-square statistic of $\chi_\nu$ = 1.07 for this fit. 

We then fit the spectrophotometric light curves to retrieve the transmission and spot contrast spectrum on a resolving power grid of R = 50. The orbital parameters ($T_\textrm{0}$, $b$, $a/R_\textrm{*}$), and the spot position and radius are fixed to the set of parameter values with the highest likelihood of the broadband light-curve fit. We fit the remaining parameters: the transit depth, the spot contrast, $u_1$, $u_2$, the two-parameter linear slope, and the scalar jitter term, with 500 live points. This transmission spectrum is shown in the middle panel of the Extended Data Fig. \ref{fig:spot-crossing}. Finally, we constrain the temperature of the occulted spot by fitting PHOENIX stellar atmosphere model spectra \citep{husser_new_2013} to the spot contrast spectrum \citep{fournier-tondreau_near-infrared_2024}. We fit the spot temperature with a wide, uninformative prior using the \texttt{emcee} MCMC package \citep{foreman-mackey_emcee_2013}. We fix the photosphere temperature to T$_*$ = 2960\,K\citep{peterson_temperate_2023} and the stellar surface gravity to $\log{g}_*$ = 5.115 dex\citep{peterson_temperate_2023}. We used four walkers and ran the fit for 5000 steps well past
the convergence of the MCMC chain (given by the Gelman-Rubin convergence test metric). We find that our most likely spot solution is 45 $\pm$ 2\,K colder than the photosphere.
We experimented with retrieving a different surface gravity for the spot compared to the photosphere \citep{fournier-tondreau_near-infrared_2024} but found this unnecessary.

The transmission spectra retrieved using both treatments of the spot-crossing event (the Gaussian model or \texttt{spotrod}) are in strong agreement (Extended Data Fig. \ref{fig:spot-crossing}). The comparison is performed on two spectra fitted at R = 50 using the $(u_1,u_2)$ limb-darkening parameterization. Outside the 1-2 $\mu$m region where saturation occurs, virtually all spectroscopic transit depths agree well within 1$\sigma$, with no important systematic deviations. This confirms that the Gaussian spot model successfully models the effect of the spot-crossing event in the transit light curves and leads to the same transmission spectrum, making our atmosphere analysis independent of the spot-crossing treatment. 

\subsection{SCARLET atmosphere retrievals}
We analyse the transmission spectrum of LP~791-18\,c by performing multiple atmosphere retrievals using the SCARLET framework \citep{benneke_strict_2015, benneke_water_2019, benneke_sub-neptune_2019, roy_is_2022,coulombe_broadband_2023, roy_water_2023, benneke_jwst_2024, piaulet-ghorayeb_jwstniriss_2024}. SCARLET performs Bayesian analyses in order to constrain the atmospheric properties (chemical abundances, cloud properties, temperature, etc.). While we test multiple retrievals on our observed spectrum, using multiple parameterizations for the atmosphere models (described below), the SCARLET retrieval method works as follows. For each set of parameters, SCARLET generates a model of the atmosphere of LP~791-18\,c in hydrostatic equilibrium (the clouds and molecular abundance modelling depend on the parameterization). It then computes the opacities associated with the molecules and clouds present in the atmosphere model, computes the associated model transit spectrum, convolves it to the bins and resolution of the observed spectrum, and performs the likelihood evaluation. For each set of parameters, the radius at the 10\,mbar reference pressure is fitted to minimize the $\chi^2$ with the observed spectrum (this includes iterating the hydrostatic equilibrium and radiative transfer computations for each radius). The atmosphere models are generated with 40 vertical (pressure) layers, and the resolution of the model is 16,000. All retrievals are performed using the nestle implementation of Bayesian nested sampling \citep{skilling_nested_2004, skilling_nested_2006, feroz_multinest_2009, mukherjee_nested_2006, shaw_efficient_2007} in order to obtain both the posterior probability distribution of the parameters studied and the Bayesian evidence of the different models tested.

\subsection{Cloud parameterizations} Given the particular shape and muted features in the transit spectrum of LP~791-18\,c, we expect cloud opacities to play a major role in shaping the observed spectrum. We thus consider multiple cloud parameterizations in our models: grey clouds parameterized by the cloud top pressure $p_{\mathrm{cloud}}$ \citep{roy_water_2023}, hazes parameterized by the Rayleigh haze enhancement factor $c_{\mathrm{haze}}$ \citep{benneke_jwst_2024}, and Mie scattering clouds parameterized by $R_{\mathrm{part}}$ (cloud particle radius in $\mu$m), $p_{\tau = 1}$ (pressure in pa where the cloud optical depth is 1), and $H_{\mathrm{cloud}}$ (cloud relative scale height) following the implementation of ref.\citep{benneke_sub-neptune_2019}. We also tested models using the sigmoid cloud parameterization \citep{constantinou_vira_2024}, or adding a cloud covering fraction \citep{welbanks_degeneracies_2019}, but find that they do not add significant insights compared to the retrievals with the clouds presented above. For the Mie scattering clouds, we use the properties of KCl particles to model the opacity profile. No matter the parameterization used for the chemistry, we find that the grey clouds are unable to generate the fading opacity profile that is present in the spectrum (which the hazes and Mie clouds can reproduce). We find that the transmission spectrum is better explained when both hazes and Mie clouds are considered in our models, with the hazes reproducing the sharp opacity rise short-wise of 1\,$\mu$m, and the Mie clouds reproducing the fading opacity of the cloud past 2\,$\mu$m. In order to avoid the Mie clouds to model the same short-wavelength opacities as the hazes (and create unnecessary degeneracies in the parameter space), we use the following log-uniform priors on the parameters: $\log c_{\mathrm{haze}} \in [-10, 8]$, $\log R_{\mathrm{part}} \in [-0.15, 0.04]$, $\log p_{\tau = 1} \in [-2, 6.7]$, $\log H_{\mathrm{cloud}} \in [0.7, 1.3]$. Without those priors, the Mie clouds are allowed to find a solution for small particle radii ($<$1\,$\mu$m) and effectively become hazes. However, in that scenario, it cannot offer a "slowly-fading" opacity continuum at the longer wavelengths. We thus use this combination of hazes and larger-particles Mie scattering clouds for the different chemistry investigations described below.

\subsection{Free chemistry retrievals} We first perform a series of free chemistry retrievals in order to assess the presence or absence of numerous molecules in LP~791-18\,c's atmosphere. In this suite of retrievals, the molecular abundances (in logarithmic space) are directly fitted as free parameters with uniform priors on their abundances (ranging from 10$^{-10}$ to 1 in volume mixing ratio). The molecules we consider in our free chemistry retrievals are CH$_4$\citep{hargreaves_accurate_2020}, H$_2$O\citep{polyansky_exomol_2018}, CO$_2$\citep{yurchenko_exomol_2020}, CO\citep{hargreaves_spectroscopic_2019}, NH$_3$\citep{coles_exomol_2019}, H$_2$S\citep{azzam_exomol_2016}, and SO$_2$\citep{underwood_exomol_2016}; whereas H$_2$ and He are modelled as filling gases (according to the Jupiter abundance ratio H$_2$/He=0.157). We also fit for the photospheric temperature of the atmosphere T$_{\mathrm{atm}}$ using a Gaussian prior (355$\pm$50\,K), with the width of the Gaussian much larger than the actual uncertainty on the equilibrium temperature to ensure that our models are marginalised over a large range of plausible temperature values. We use the hazes and Mie scattering clouds described above to model the impact of aerosols on the spectrum. We find that methane is the only detected molecule in our retrieval, along with hazes. We obtain upper bounds on the abundances of the other molecules (Extended Data Table \ref{tab:mol_constr}). By performing retrievals where we take out one molecule at a time and compare the evidence in a Bayesian model comparison scheme, we find that methane is detected at 4.47\,$\sigma$ significance, and that hazes are detected at the 5.36$\sigma$ level (Extended Data Table \ref{tab:detect2025}). Our inferences on the molecular abundances are shown in Extended Data Table \ref{tab:mol_constr}. Our retrievals constrain the methane abundance to above the 0.1\% level, and display a bimodal distribution which stems from a subtle degeneracy with the amount of hazes in the models.

From the free chemistry retrieval including all molecules, we compute our inference on the CH$_4$ to CO$_2$ abundance ratio. CH$_4$/CO$_2$ is a useful quantity when it comes to comparisons to other characterized low-temperature sub-Neptunes, and to linking the upper atmosphere composition to interior processes \citep{madhusudhan_carbon-bearing_2023, benneke_jwst_2024} and formation scenarios for exoplanets \cite{yang_chemical_2024}. We find that the CH$_4$/CO$_2$ ratio in LP~791-18\,c's atmosphere is greater than 12 at 2$\sigma$ confidence. We further estimate the mean molecular weight of LP~791-18\,c's atmosphere from our molecular abundance samples, and find a constraint of $7.68 ^{+2.24} _{-1.48}$ atomic mass units (Fig. \ref{fig:measurements}, Extended Data Table \ref{tab:mol_constr}).

Using centred-log-ratio priors on the molecular abundance parameters does not change the conclusions of our free chemistry analysis of LP~791-18\,c's transit spectrum. In order to ensure that the choice of H$_2$/He background gas does not affect the conclusions of our analysis, we run an atmosphere retrieval where we fit for the CLR-transformed abundances of the same molecules (with the only difference being that now H$_2$ alone acts as the unparameterized gas) \citep{benneke_atmospheric_2012}. The main difference on the retrieved posteriors is that the lower bound retrieved on the only detected molecule, methane, is much stronger than in the usual log-uniform prior case with H$_2$/He as the background gases. This also has the effect of bringing the derived mean molecular weight around $\mu$ = 16 amu, the expected value for a pure methane gas. This is not surprising, as the same behaviour was noticed in the recent analysis of GJ 9827\,d's transit spectrum, in which the CLR retrieval was jumping to pure water models, with water being the only detected molecule in that instance\citep{piaulet-ghorayeb_jwstniriss_2024}. Despite this difference, the conclusions on the atmosphere composition remain broadly consistent.

\subsection{Chemically consistent retrievals} We perform another series of retrievals in which we model the molecular abundances at chemical equilibrium. In that parameterization, the metallicity and carbon-to-oxygen ratio (C/O) are the fitted parameters, and the molecular abundances are calculated at each step in order to have the gas at chemical equilibrium for every temperature-pressure layer of the atmosphere model. The following molecules are considered for this chemically consistent parameterization: H$_2$, He, H$_2$O, CH$_4$, CO, CO$_2$, NH$_3$, HCN, N$_2$, SO$_2$, H$_2$S, CS$_2$, OH.  The temperature and cloud parameterizations are unchanged from the free chemistry analyses described above. The uniform priors we use on the chemical equilibrium parameters are the following: $\log M \in [0, 4]$, C/O $\in [0, 3]$. In order to account for the potential presence of H$_2$S in LP~791-18\,c's atmosphere (which shows a mode in the posterior of the free retrieval, and which could help produce a particular 2.8\,$\mu$m feature, see below), we perform a new retrieval with the same parameters, but for which we additionally fit for a vertically-constant abundance of H$_2$S in the atmosphere \citep{tsai_photochemically_2023,piaulet-ghorayeb_jwstniriss_2024}. For each sampled abundance of H$_2$S, we scale the abundances of the other gases obtained from the chemical equilibrium computation to preserve their abundance ratios.  

We find that both versions of the chemically consistent retrievals are in agreement, giving metallicity measurements of 316$^{+82}_{-70}$ in the standard case and 309$^{+89}_{-69}$ in the cases with added H$_2$S, which in the end does not help the fit in terms of the Bayesian evidence. In all cases, the C/O remains largely unconstrained as no oxygen-bearing species is strongly detected, and since CH$_4$ remains the dominant opacity source over much of the metallicity-C/O parameter space. We can also derive a mean molecular weight measurement from the samples of the retrievals, and find that they are broadly consistent with the free chemistry retrieval inferences (Extended Data Table \ref{tab:mol_constr}).

In summary, while some differences exist in the posteriors obtained from the different retrievals tested in this study, they all depict a consistent picture of a hazy atmosphere with methane features peaking through the aerosols. 

\subsection{On the unknown 2.8\,$\mu$m feature}
We find that the transmission spectrum of LP~791-18\,c displays what looks like an absorption feature peaking around 2.8\,$\mu$m; and that feature is present in the spectrum obtained from both our reduction pipelines. While this feature can be partially fitted by the addition of H$_2$S, H$_2$O, or NH$_3$ in our retrievals, none of those molecules represent a truly good fit, explaining why we do not robustly detect those molecules, and why we obtain weak constraints on their abundances. In order to try and explain the presence of that absorption band, we inspect the absorption cross-sections of the HITRAN and of the ExoMol databases. We find that the only molecules that show some significant opacity in a narrow band around 2.8\,$\mu$m and whose opacity profile could remain consistent with the rest of the transmission spectrum over the NIRSpec/PRISM wavelength range are OH, HO2, and HOCl. While OH comes as an obvious product of water photodissociation, we find that photochemical modelling of temperate sub-Neptunes can only predict negligible amounts of OH in those atmospheres \citep{yang_chemical_2024}. We run free chemistry retrievals including OH and find that it would have to be abundant at the percent level, orders of magnitude above the photochemical predictions, in order to explain the observed feature. Similarly, we do not find compelling evidence that HOCl and HO$_2$ are expected photochemical products that would be present in observable amounts in sub-Neptune atmospheres. We thus conclude that the feature could be due to an unknown and unexpected molecular absorber we did not consider here. It could also be that it is an absorption feature produced by a specific aerosol species, but the study of all possible cloud species and their respective absorption and scattering cross-sections is outside the scope of this work. Finally, the feature could also be stochastic, as it is 1-2\,$\sigma$ away from models with H$_2$S/NH$_3$/H$_2$O. 



\subsection{Stellar contamination}
During exoplanet transits, unocculted stellar spots and faculae can potentially mimic absorption features from the planet's atmosphere via the transit light-source (TLS \citep{rackham_transit_2018, rackham_transit_2019}) effect. The TLS contamination level is expected to be more important for active (significant spots and faculae coverage) and cold (more molecules can be stable in the star's photosphere) stars such as TRAPPIST-1 \citep{lim_atmospheric_2023, radica_promise_2025}, or LP 791-18. Furthermore, some signs of stellar activity were observed in the spot-crossing event that occurred during the transit of LP~791-18\,c. Hence, we investigate the potential stellar contamination of our observed spectrum in order to see whether it affects our results in any meaningful way.

To start, methane is the only detected molecule in the transmission spectrum of LP~791-18\,c, and this molecule is only chemically stable at temperatures much cooler than that of stellar photospheres. Therefore, it is impossible for this methane absorption signal to be produced by the TLS effect, and since it is our only strongly detected species, it must be a constituent of the planetary atmosphere. This is in contrast to water or carbon monoxide, which are molecules that can be present at the hot temperatures of late M dwarfs, and thus can mimic atmospheric absorption from the exoplanet. However, no features from such molecules are observed in our transmission spectrum.

Overall slanted trends can also be introduced into transmission spectra by TLS effects \citep{fournier-tondreau_near-infrared_2024}, which could mimic some cloud features. Hence, to quantify and evaluate the impact TLS could have on our observed spectrum and our atmosphere inference, we test our atmosphere retrievals by jointly fitting for stellar contamination. That is, we perform additional versions of our free chemistry and chemically consistent atmosphere retrievals in which we jointly fit for the TLS contamination following the methodology described in previous studies \citep{fournier-tondreau_near-infrared_2024, piaulet-ghorayeb_jwstniriss_2024}. We thus add to our fits three parameters that describe the stellar heterogeneities: the spot covering fraction $f_{\mathrm{spot}} \in [0,0.5]$, the spot temperature difference $\Delta T_{\mathrm{spot}} \in [-600, 0]$, and the photosphere temperature $T_{\mathrm{phot}}$ for which we use a Gaussian prior based on the measured stellar temperature (2960 $\pm$ 55\,K \citep{peterson_temperate_2023}). We choose to fit for stellar spots since their presence could produce the rise in transit depths at short wavelengths observed in our transmission spectrum \citep{moran_high_2023}.


We find that the retrievals are consistent with no spots ($f_{\mathrm{spot}} < 0.07$ at $ 2 \sigma)$, and that the inferred results are mostly unaffected by the addition of the TLS parameters (Fig. \ref{fig:measurements}, Extended Data Fig. \ref{fig:wellMixedCorner}, Extended Data Fig. \ref{tab:mol_constr}). The retrieved spots parameters, in terms of the covering fraction ($<7\%$ at 2$\sigma$) and temperature contrast ($<615\,$K, a parameter highly unconstrained when the spot covering fraction approaches zero) are consistent with the stellar activity level derived from the spotrod analysis (3.6\% covering fraction and 35\,K contrast). While the spots can explain in part the rising trend in the transit depths towards short wavelengths, it seems unable to reproduce the full amplitude of the signal and the muted nature of the transmission spectrum, and so the detection of hazes stays robust (3.54$\sigma$) when adding stellar spots to the fit (Extended Data Table \ref{tab:detect2025}). We further find that adding TLS hurts the evidence of the model, and we thus conclude that the hazes and methane signals come from the planetary atmosphere.

\subsection{Miscible envelope models}
In order to investigate the miscible envelope sub-Neptune scenario \citep{benneke_jwst_2024} in the context of LP~791-18\,c, and to explore under which conditions vertical mixing could enrich the upper atmosphere in CO$_2$, we produce a grid of radiative SCARLET forward models which we couple with the VULCAN chemical kinetics framework \citep{tsai_vulcan_2017}. The SCARLET forward models are all produced assuming a solar C/O, an internal temperature of 30\,K and full heat redistribution. The parameters we vary are the metallicity (from solar to 1000 $\times$ solar), and the Bond albedo (0.0, 0.2, 0.3; Extended Data Fig. \ref{fig:miscibleEnv}). For each set of parameters, SCARLET produces a forward atmosphere model in hydrostatic equilibrium (as it was described in the retrieval section), but this time around the model is made of 60 pressure layers, and the temperature-pressure profile is solved so that the atmosphere is in thermo-chemical and radiative equilibrium. Once the SCARLET model is converged, the temperature-pressure profile is used as input for a VULCAN chemical kinetics calculation (with the CNOH network) using the same metallicity and C/O, and using a K$_{zz}$ of 10$^4$\,cm$^2$/s to model the vertical mixing in the atmosphere \citep{benneke_jwst_2024}. Then, for each resulting model, we extract the CH$_4$/CO$_2$ ratio at 10\,mbar, and display that value in Extended Data Fig. \ref{fig:miscibleEnv}. We further show the mixing ratios and temperature structure of a sample model from our grid in Extended Data Fig. \ref{fig:miscibleEnv}.

\subsection{Population trends}
We investigate trends in the atmospheric signal strengths measured in transmission spectroscopy across the sub-Neptune population. We follow the procedure described in ref. \citep{brande_clouds_2024} to evaluate a signal amplitude strength for the 1.4 $\mu$m H$_2$O/CH$_4$ absorption band in units of scale heights (assuming $\mu$=3.05: H$_{\mu=3.05}$ as in the HST study) for the targets recently characterized with JWST. For planets that have been observed by HST/WFC3 and that have not been observed by NIRISS SOSS (GJ 3470\,b and GJ 1214\,b), we use the same HST/WFC3-derived values presented in ref. \citep{brande_clouds_2024}. For all the other sub-Neptunes (K2-18\,b \citep{madhusudhan_carbon-bearing_2023}, TOI-270\,d \citep{benneke_jwst_2024}, LP~791-18\,c, GJ 9827\,d \citep{piaulet-ghorayeb_jwstniriss_2024}, GJ 3090\,b, Ahrer et al., under review.) we run atmosphere retrievals on the JWST transmission spectra and measure the 1.4\,$\mu$m absorption signal from the retrieval samples. Our retrievals are performed using the free chemistry parameterization, allow for hazes as well as grey clouds, and otherwise follow the methods described in this manuscript. We use 300 samples binned at a resolving power of 300 to estimate the distribution of signal amplitudes in units of scale heights for each planet. For TOI-270\,d, GJ~9827\,d and GJ~3090\,b, we post-process the exact retrievals presented in the respective papers to evaluate the signal strengths. For K2-18\,b, we use the transmission spectrum from ref.\citep{madhusudhan_carbon-bearing_2023}, and run two versions of the retrieval: one where we take the spectrum as is (grey point in Fig. \ref{fig:population}), and one where we fit for an offset between the NIRISS and NIRSpec parts of the spectrum. In the case of TOI-836\,c, which only has a NIRSpec/G395H 2.7-5.2\,$\mu$m spectrum, we still evaluate the 1.4\,$\mu$m absorption band from the samples. Since the spectrum is featureless over the NIRSpec/G395H band, we retrieve a small 1.4\,$\mu$m predicted signal. This is expected since H$_2$O and CH$_4$ absorption bands are larger over the G395H wavelength range than at 1.4\,$\mu$m. Finally, for TOI-421\,b, the amplitude of the 1.4\,$\mu$m feature was already measured in ref.\citep{davenport_toi-421_2025} and we use that measurement. 

We produce an updated version of the figure investigating atmospheric signal strengths versus the equilibrium temperature of each sub-Neptune (for full heat redistribution and 0.3 Bond albedo, Fig. \ref{fig:population}) for the planets described above. We further color the planets as a function of the atmospheric inference. For instance, atmospheres that were observed to be mostly free of cloud opacity (allowing strong detections of all molecular bands, such as on K2-18\,b \citep{madhusudhan_carbon-bearing_2023}, TOI-270\,d \citep{benneke_jwst_2024}, GJ 9827\,d \citep{piaulet-ghorayeb_jwstniriss_2024}, GJ 3090\,b (Ahrer et al., under review), and TOI-421\,b \citep{davenport_toi-421_2025}) are displayed in yellow. Atmospheres for which important clouds or hazes opacity were robustly observed are shown in red (LP~791-18\,c, GJ 1214\,b \citep{gao_hazy_2023}, and GJ 3470\,b \citep{beatty_sulfur_2024}). For TOI-836\,c, it is still unclear whether the attenuated spectrum is due to clouds or to high mean molecular weight atmospheres, we thus show it in orange.  

We also display the sub-Neptune absorption signal amplitudes as a function of the incoming high-energy irradiation. We estimate that quantity by computing the stellar X-ray-to-UV flux following refs. \citep{jackson_coronal_2012, owen_evaporation_2017}. This method uses the stellar mass along with the stellar age to estimate the current-day luminosity of the star, which we convert into the incoming flux on the planets. Given the large uncertainties on the stellar ages, our estimations of the X-ray-to-UV fluxes have substantial uncertainties as well. We show that population in Fig. \ref{fig:population}.



\end{methods}


\clearpage
\begin{addendum}

 \item[Data Availability] 
 The data used in this work are publicly available in the Mikulski Archive for Space Telescopes ({\small \url{https://archive.stsci.edu/}}) under GTO programme 1201 (principal investigator D.L.).
 
 \item[Code Availability] 
 This research made use of the Astropy\citep{astropy:2013, astropy:2018, astropy:2022}, Matplotlib\citep{Hunter:2007}, NumPy\citep{harris2020array} and SciPy\citep{2020SciPy-NMeth} Python packages. The open-source codes that were used throughout this work are listed below:\\
\texttt{batman} (\url{https://github.com/lkreidberg/batman});\\
\texttt{emcee} (\url{https://emcee.readthedocs.io/en/stable/});\\
\texttt{nestle} (\url{http://kylebarbary.com/nestle/});\\
\texttt{Eureka!} (\url{https://eurekadocs.readthedocs.io/en/latest/});\\
\texttt{ExoTEDRF} (\url{https://exotedrf.readthedocs.io/en/latest/});\\
\texttt{VULCAN} (\url{https://github.com/exoclime/VULCAN});\\
 
 \item[Acknowledgements]

 P.-A.R. and B.B acknowledge financial support from the Natural Sciences and Engineering Research Council (NSERC) of Canada and from the Canadian Space Agency under grant 23JWGO2A05. P.-A.R. and L.-P.C. further acknowledge support from the University of Montreal and the Trottier Institute for Research on Exoplanets (IREx). B.B. also acknowledges financial support from the Fond de Recherche Qu\'eb\'ecois-Nature et Technologie (FRQNT; Qu\'ebec). M.F.T. acknowledges financial support from the Clarendon Fund Scholarship and the Fonds de Recherche du Qu\'ebec--Nature et technologies (FRQNT). R.A. acknowledges the Swiss National Science Foundation (SNSF) support under the Post-Doc Mobility grant P500PT\_222212 and the support of the Institut Trottier de Recherche sur les Exoplane\`etes (IREx). N.B.C. acknowledges support from an NSERC Discovery Grant, a Tier 2 Canada Research Chair, and an Arthur B.\ McDonald Fellowship. The authors also thank the Trottier Space Institute and the Trottier Institute for Reasearch on Exoplanets for their financial support and dynamic intellectual environment. L.D. is a Banting and Trottier Postdoctoral Fellow and acknowledges support from the Natural Sciences and Engineering Research Council (NSERC) and the Trottier Family Foundation. D.J.\ is supported by NRC Canada and by an NSERC Discovery Grant. This project has been carried out within the framework of the National Centre of Competence in Research PlanetS supported by the Swiss National Science Foundation under grant 51NF40\_205606. S.P.\ acknowledges the financial support of the SNSF. R.J.M. is supported by NASA through the NASA Hubble Fellowship grant HST-HF2-51513.001, also awarded by the Space Telescope Science Institute, which is operated by the Association of Universities for Research in Astronomy, Inc., for NASA, under contract NAS 5-26555. J.T. acknowledges funding support by the TESS Guest Investigator Program G06165.

\item[Author Contributions] 
P.-A.R. and B.B. led the writing of this manuscript. P.-A.R. carried out the data reduction. P.-A.R. performed the light curve fitting with contributions from M.F.-T.. The atmospheric analyses were performed by P.-A.R.. All co-authors provided significant comments and suggestions to the manuscript.

\item[Competing Interests] The authors declare no competing interests.
 
\item[Correspondence] Correspondence and requests for any materials presented in this work should be addressed to Pierre-Alexis Roy.~(email: pierre-alexis.roy@umontreal.ca).

\end{addendum}
 \clearpage

\clearpage

\begin{supplementary}
\setcounter{figure}{0}
\renewcommand{\figurename}{Extended Data Figure}
\renewcommand{\tablename}{Extended Data Table}

\clearpage

\begin{figure*}[t!]
\begin{center}
\includegraphics[width=0.9\linewidth]{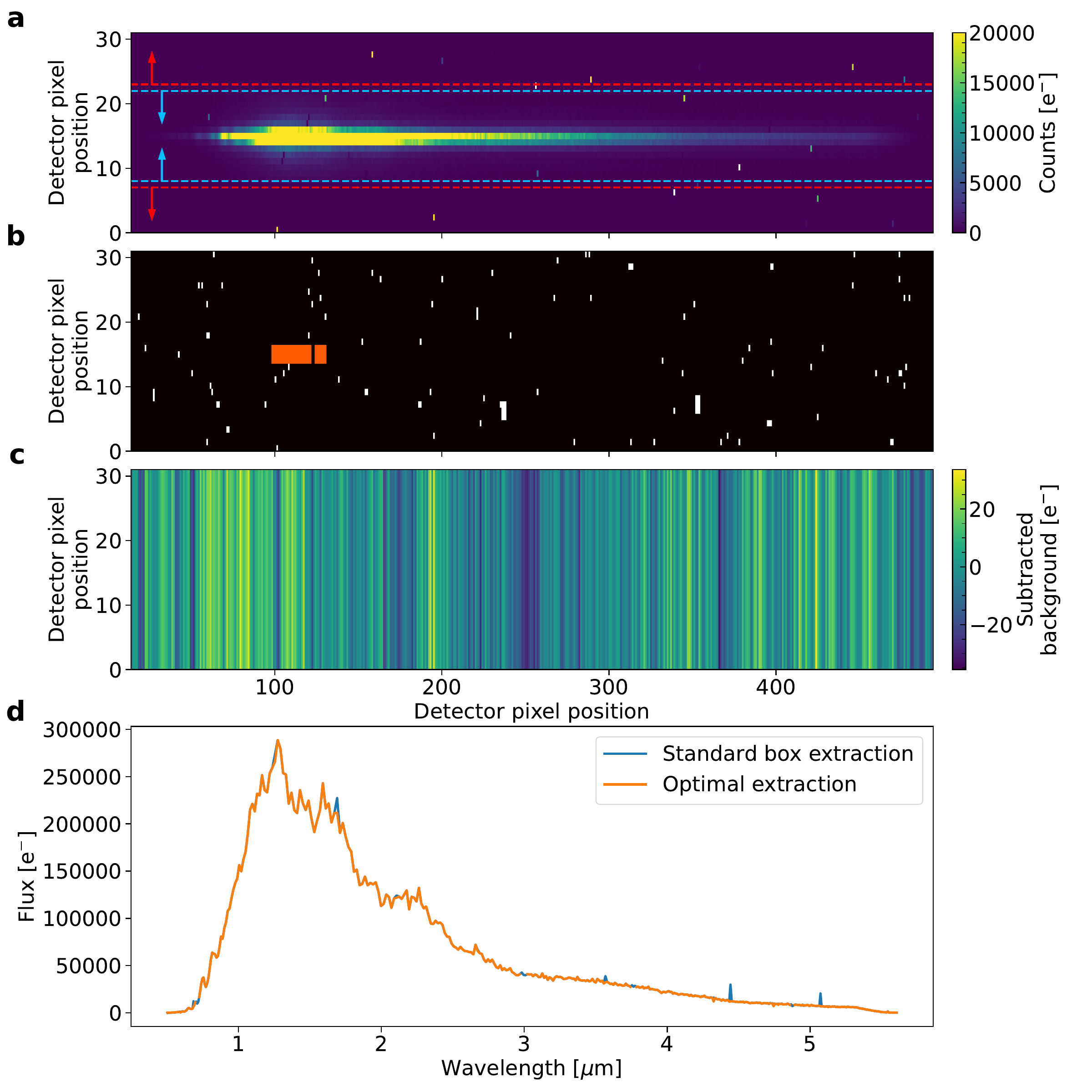}
\end{center}
\caption{\textbf{Extraction of the stellar spectrum for a sample integration.} \textbf{a.} Sample detector frame after ramp-fitting and calibration (Stage 2). The blue dashed lines highlight the aperture for the spectral extraction, whereas the red dashed lines highlight the pixels used for the column-by-column background subtraction. The color scale is cut at 20000 electrons so that more of the trace is visible. \textbf{b.} Data quality map of the frame above, with saturated pixels in orange and other bad pixels in white. \textbf{c.} Subtracted background from the 1/$f$ column-by-column correction at the integration level. \textbf{d.} Stellar spectrum of LP 791-18 extracted from the sample integration. The stellar flux is shown in electrons, has neither been flat-fielded nor throughput-corrected, and is shown for both the optimal extraction (orange) and the box extraction (blue). }
\label{fig:reduction}
\end{figure*}

\begin{figure*}[t!]
\begin{center}
\includegraphics[width=0.80\linewidth]{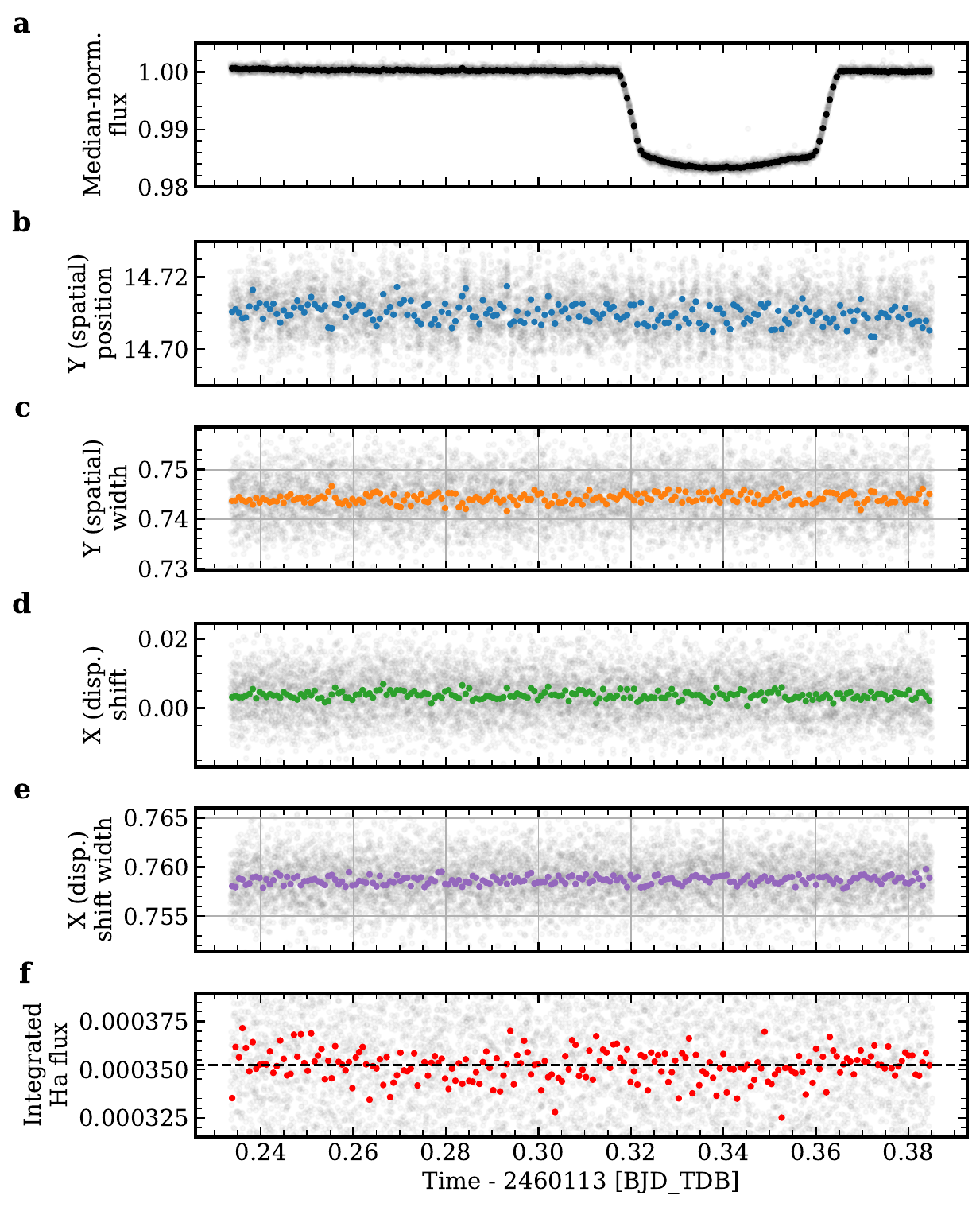}
\end{center}
\caption{\textbf{Summary diagnostics measurements for the transit observation of LP~791-18\,c.}  \textbf{a.} Median-normalized white light curve of the transit observation of LP~791-18\,c. The raw observations are shown in light grey with a binned version in black. \textbf{b.} Measured centre of the trace on the NIRSpec detector (light grey) with a binned version in blue. \textbf{c.} Measured width of the trace on the NIRSpec detector (light grey) with a binned version in orange. \textbf{d.} Measured displacement of the trace in the dispersion direction (light grey) with a binned version in green. \textbf{e.} Width of the cross-correlation peak for the measurement of the displacement in the dispersion direction (light grey) with a binned version in purple. \textbf{f.} Integrated H$\alpha$ flux over the time series (light grey) with a binned version in red. All diagnostics measurements are well-behaved.}
\label{fig:diagnostics}
\end{figure*}

\begin{table}
\caption{Stellar and planetary parameters for LP~791-18\,c used or derived in this work. }
\begin{tabular}{ccccc}
\hline
\hline
Parameter &  Units & Prior & Value &  Comment \\
\hline
\hline 
\textit{Stellar Parameters} &&&&\\
T$_\mathrm{eff}$ & K & - & $2960^{+55}_{-55}$ & P+23$^1$ \\
R$_\star$ & R$_\mathrm{sun}$ & - & $0.182^{+0.007}_{-0.007}$ & P+23\\
M$_\star$ & M$_\mathrm{sun}$ & - & $0.139^{+0.005}_{-0.005}$ & P+23\\
log$_{10}$g & dex (cgs) & - & $5.115^{+0.094}_{-0.094}$ & P+23\\
$[$Fe/H$]$ & - & - & $-0.09^{+0.19}_{-0.19}$ & P+23\\
\hline
\textit{Planet Parameters} &&&&\\
P & day & - & $4.9899093^{+0.0000074}_{-0.0000072}$ & P+23\\
T$_0$ & BJD$_\mathrm{TDB}^{\ \ \ \ \ 2}$ & $\mathcal{U}(113.24, 113.44)$ & $113.3411445^{+0.0000064}_{-0.0000063}$ & This paper\\
R$_p$/R$_\star$ & - & $\mathcal{U}(0.01, 0.223607)$ & $0.12399^{+0.00012}_{-0.00011}$ & This paper\\
b & - & $\mathcal{U}(0, 1)$ & $0.052^{+0.040}_{-0.035}$ & This paper\\
i & degrees & - & $89.920_{-0.062}^{+0.054}$ & This paper\\
a/R$_s$ & - & $\mathcal{U}(17.5, 52.5)$ & $37.091^{+0.047}_{-0.091}$ & This paper\\
a & AU & - & $0.0315^{+0.0012}_{-0.0012}$ & This paper\\
R$_p$ & R$_\oplus$ & - & $2.47^{+0.09}_{-0.09}$ & This paper\\
M$_p$ & M$_\oplus$ & - & $7.1^{+0.7}_{-0.7}$ & P+23\\
\hline
\hline
\end{tabular}
\footnotesize{\\$^1$ Peterson et al. (2023)\citep{peterson_temperate_2023}. $^2$ Times are offset by 2460000 days.}\\
\label{tab:bestfitparams} 
\end{table}

\begin{figure*}[t!]
\begin{center}
\includegraphics[width=0.9\linewidth]{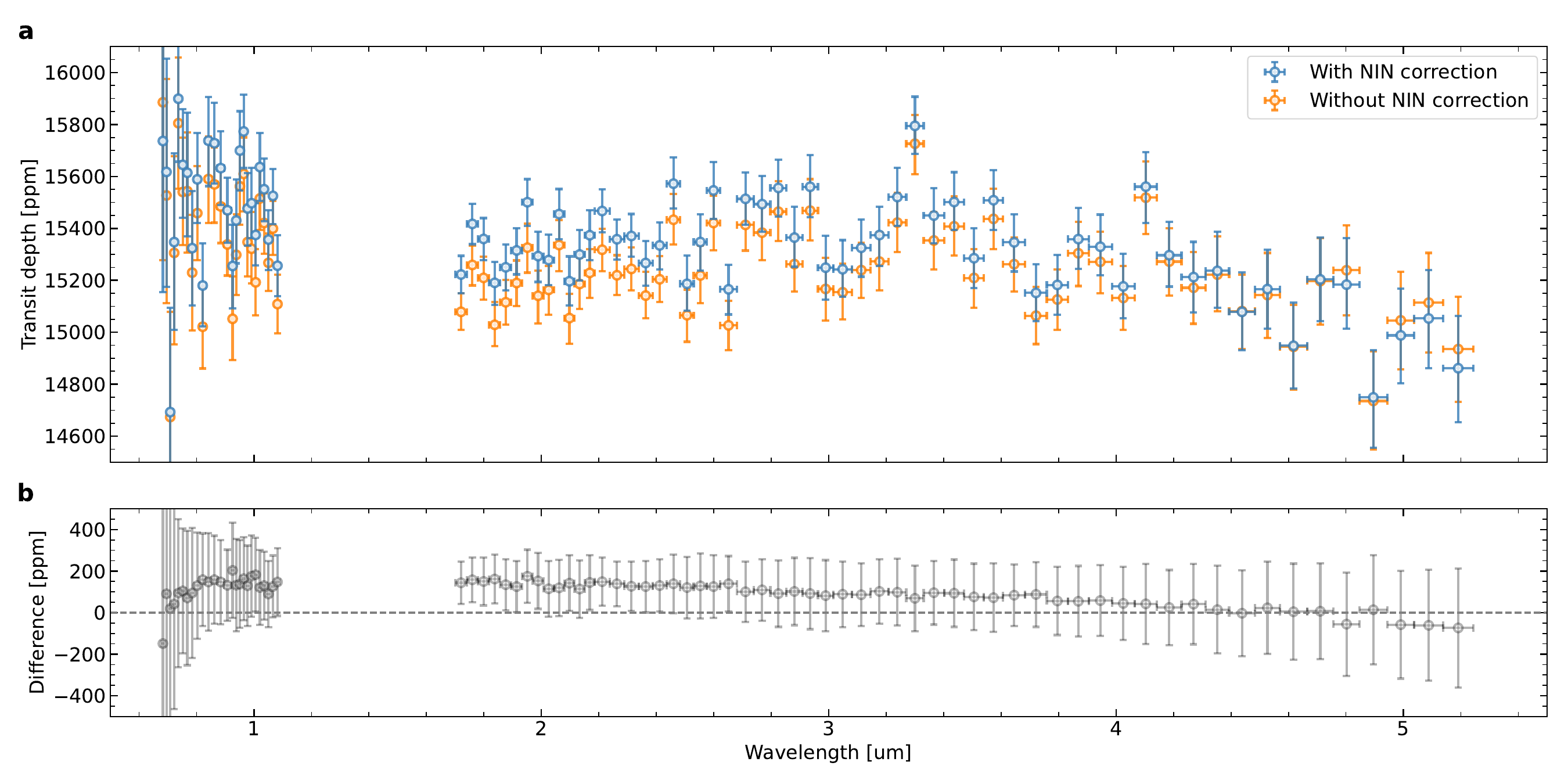}
\includegraphics[width=0.9\linewidth]{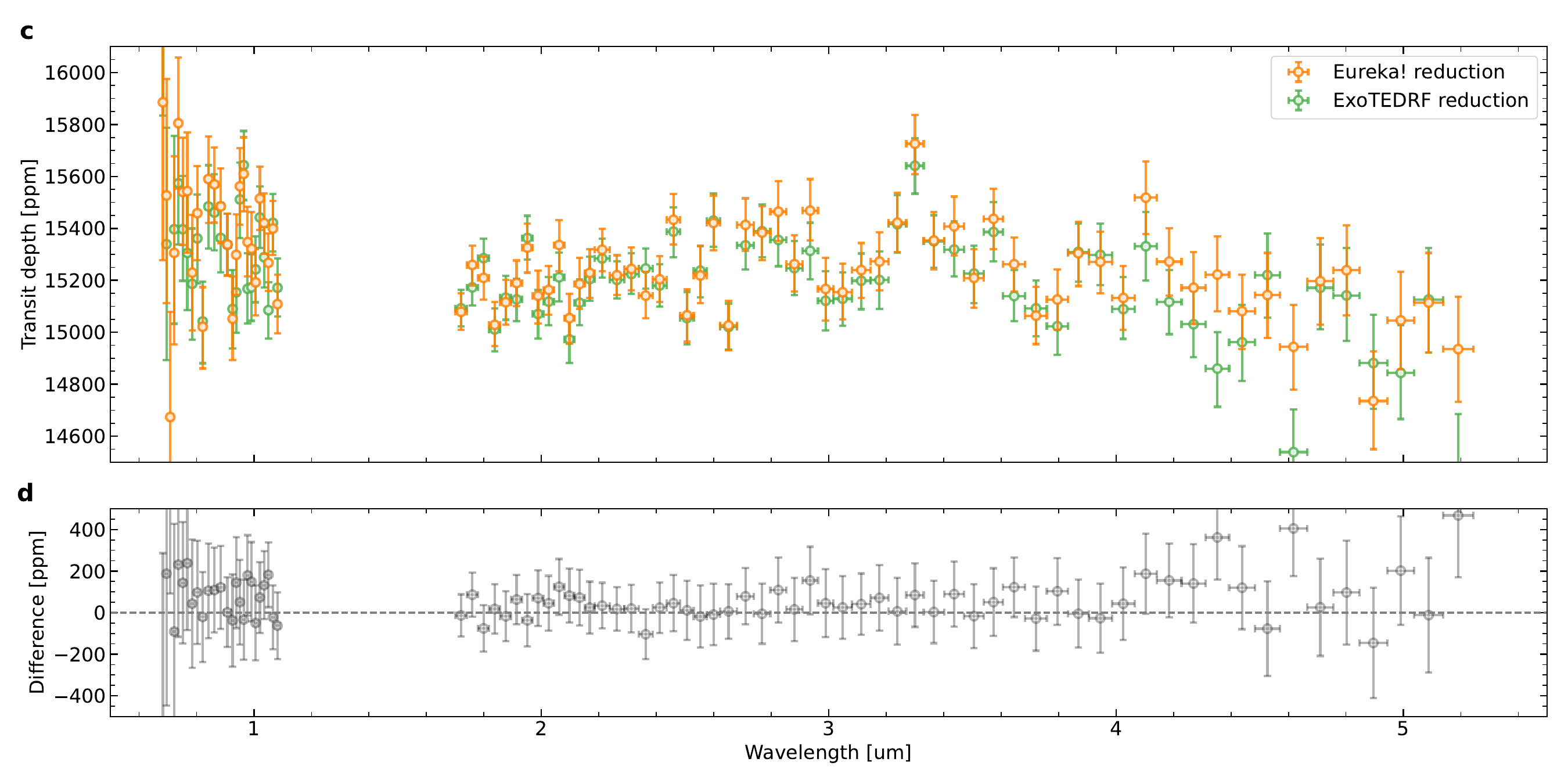}
\end{center}
\caption{\textbf{Effect of the Non-linear-illumination (NIN) correction and of the different reduction pipelines on the transmission spectrum of LP~791-18\,c.} \textbf{a.} Transmission spectra of the sub-Neptune LP~791-18\,c with 1$\sigma$ error bars obtained when running the complete analysis from a standard stage 1 reduction with (blue) and without (orange) the NIN correction step. \textbf{b.} Residuals in the transit spectrum between the reductions with and without the NIN correction. The NIN correction step rectifies a wide trend in the transmission spectrum which is most important around 2\,$\mu$m and decays towards longer wavelengths. The effect is similar to that observed in the case of TRAPPIST-1\,g (Benneke et al., under review). \textbf{c.} Transmission spectra of the sub-Neptune LP~791-18\,c with 1$\sigma$ error bars obtained when using the light curves produced by the Eureka! (orange) and ExoTEDRF (green) reductions. \textbf{d.} Residuals in the transit spectrum between the Eureka! and ExoTEDRF reductions. The NIN correction is turned off in the Eureka! reduction (to isolate the difference between the two reduction pipelines). Both transmission spectra are in agreement.}
\label{fig:NIN}
\end{figure*}


\begin{figure*}[t!]
\begin{center}
\includegraphics[align=c,width=0.3\linewidth]{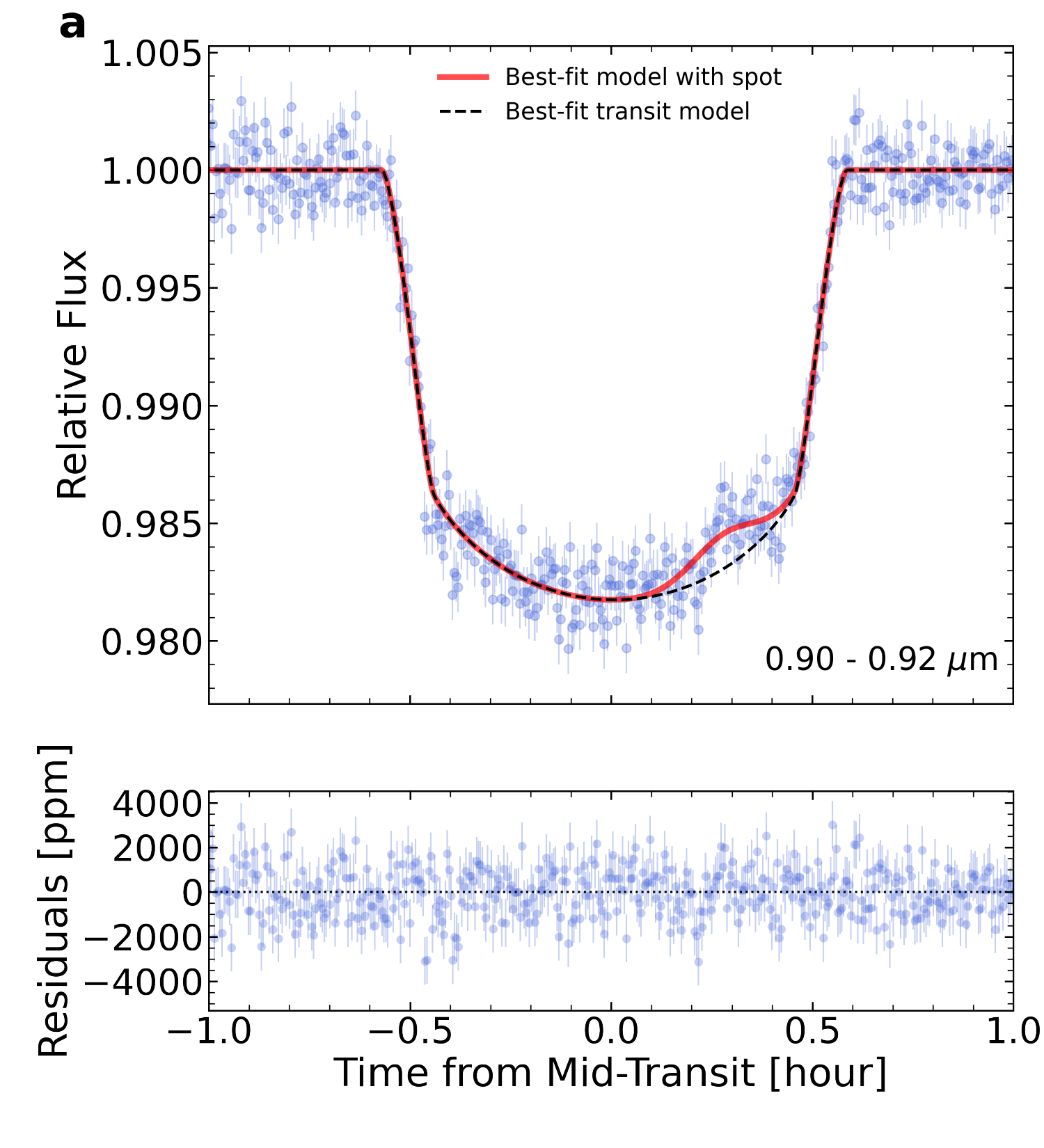}
\includegraphics[align=c,width=0.3\linewidth]{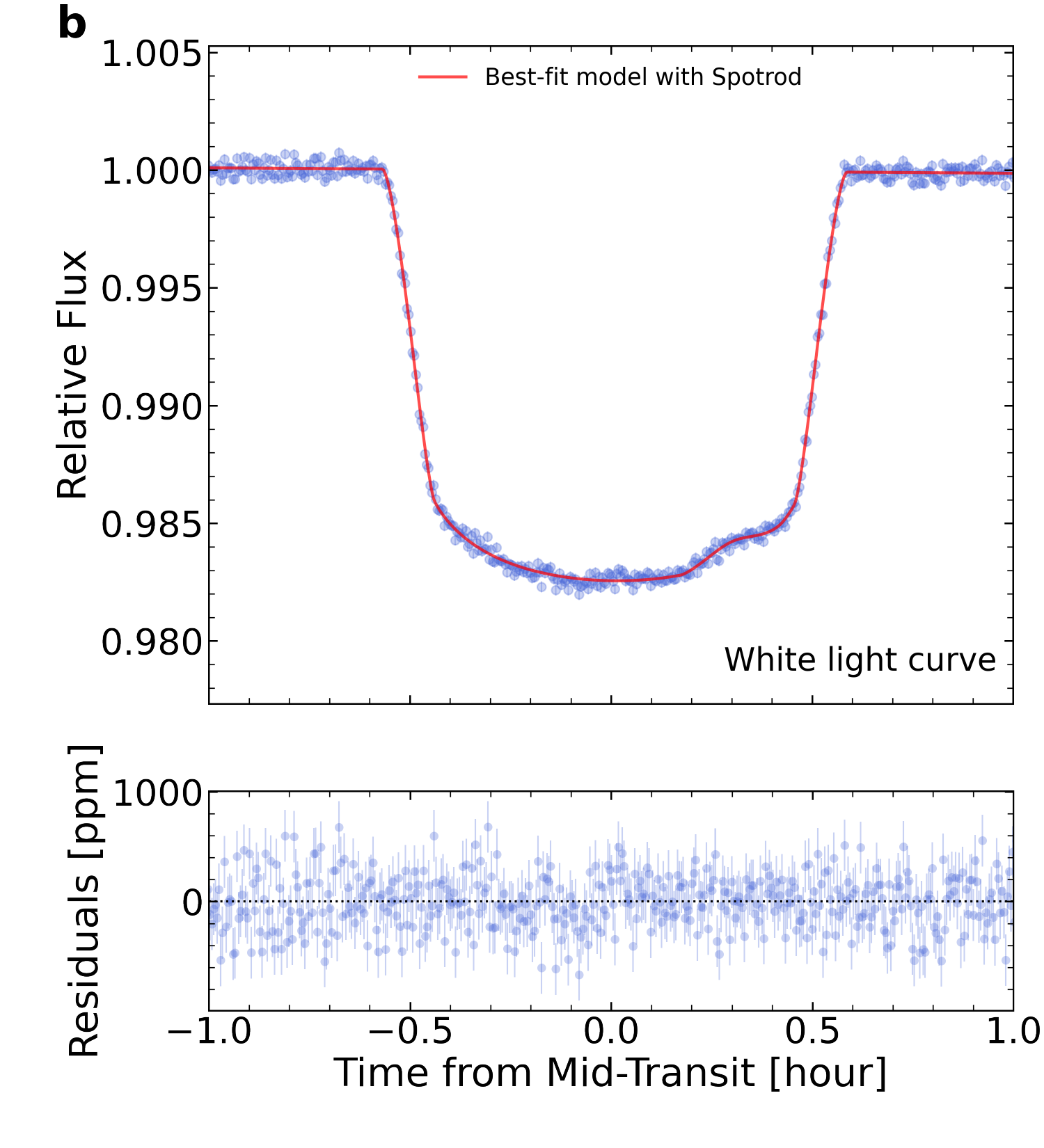}
\includegraphics[align=c,width=0.3\linewidth]{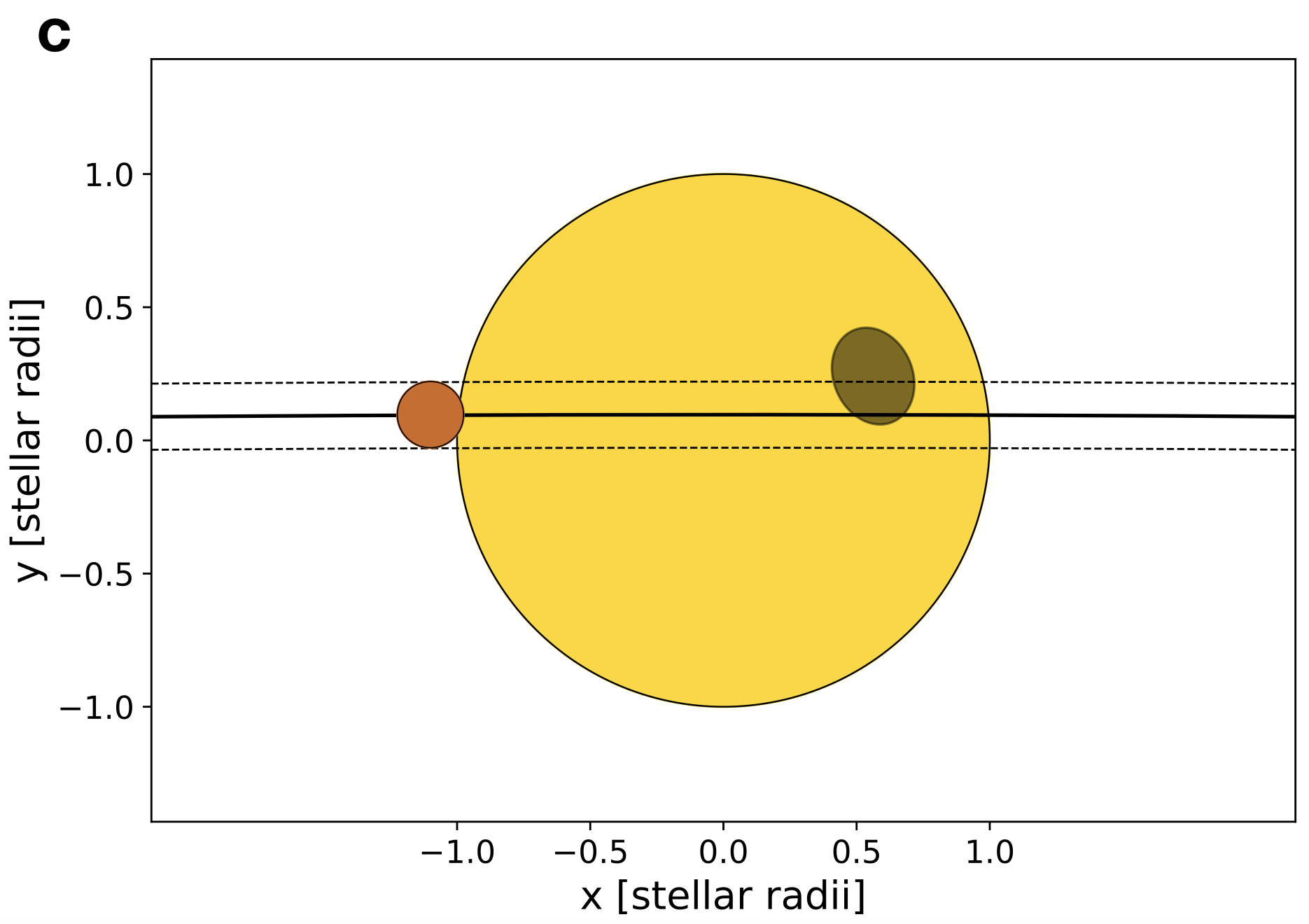}
\includegraphics[width=0.95\linewidth]{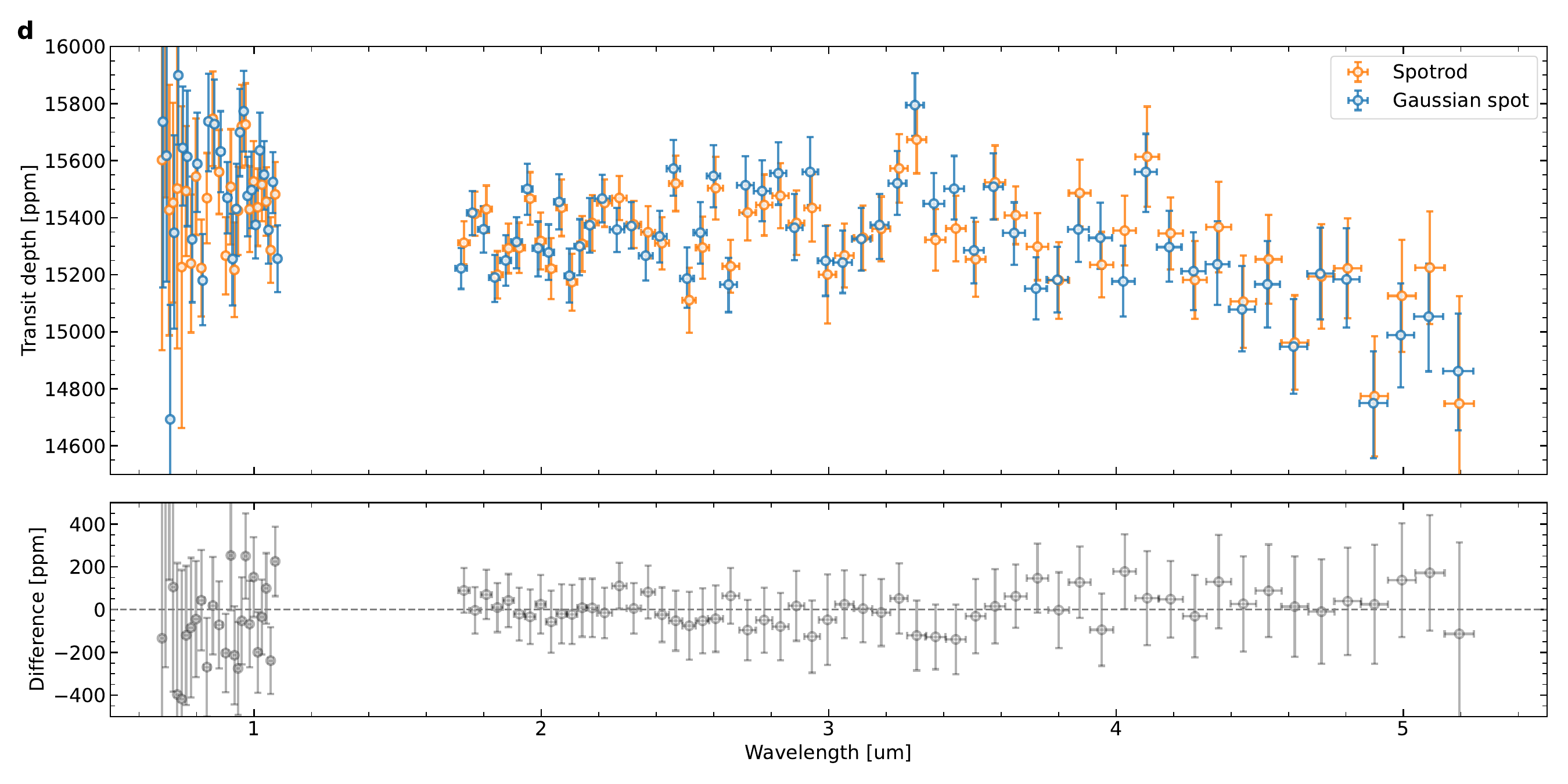}

\end{center}
\caption{\textbf{Comparison of the treatment of the spot-crossing event.} \textbf{a.} Example light-curve fit using the Gaussian spot model. The top panel shows the systematics-corrected light-curve for the 0.90-0.92\,$\mu$m bin with the best-fitting transit model with (red) and without (dotted black) the Gaussian spot model. The bottom panel shows the residuals with the best fitting model. The data is binned per 16 second increments for visual clarity. \textbf{b.} White-light-curve fit using the \texttt{spotrod} model. The top panel shows the systematics-corrected white-light curve with the best-fitting model, including the modelled spot crossing event. The bottom panel shows the residuals of the fit. Again, the data is shown in 16 s bins for clarity. \textbf{c.} Graphical representation of the spot crossing event inferred from the spotrod fit shown in b. \textbf{d.} Transmission spectra obtained from both methods using the same R=50 spectroscopic bins. Because of a slightly different set of orbital parameters used in the \texttt{Spotrod} spectroscopic fit, it is offset by 50 ppm. Both spectra are fully consistent within their displayed 1$\sigma$ error bars, and show no systematic discrepant trends.}
\label{fig:spot-crossing}
\end{figure*}

\begin{figure*}[t!]
\begin{center}
\includegraphics[width=\linewidth]{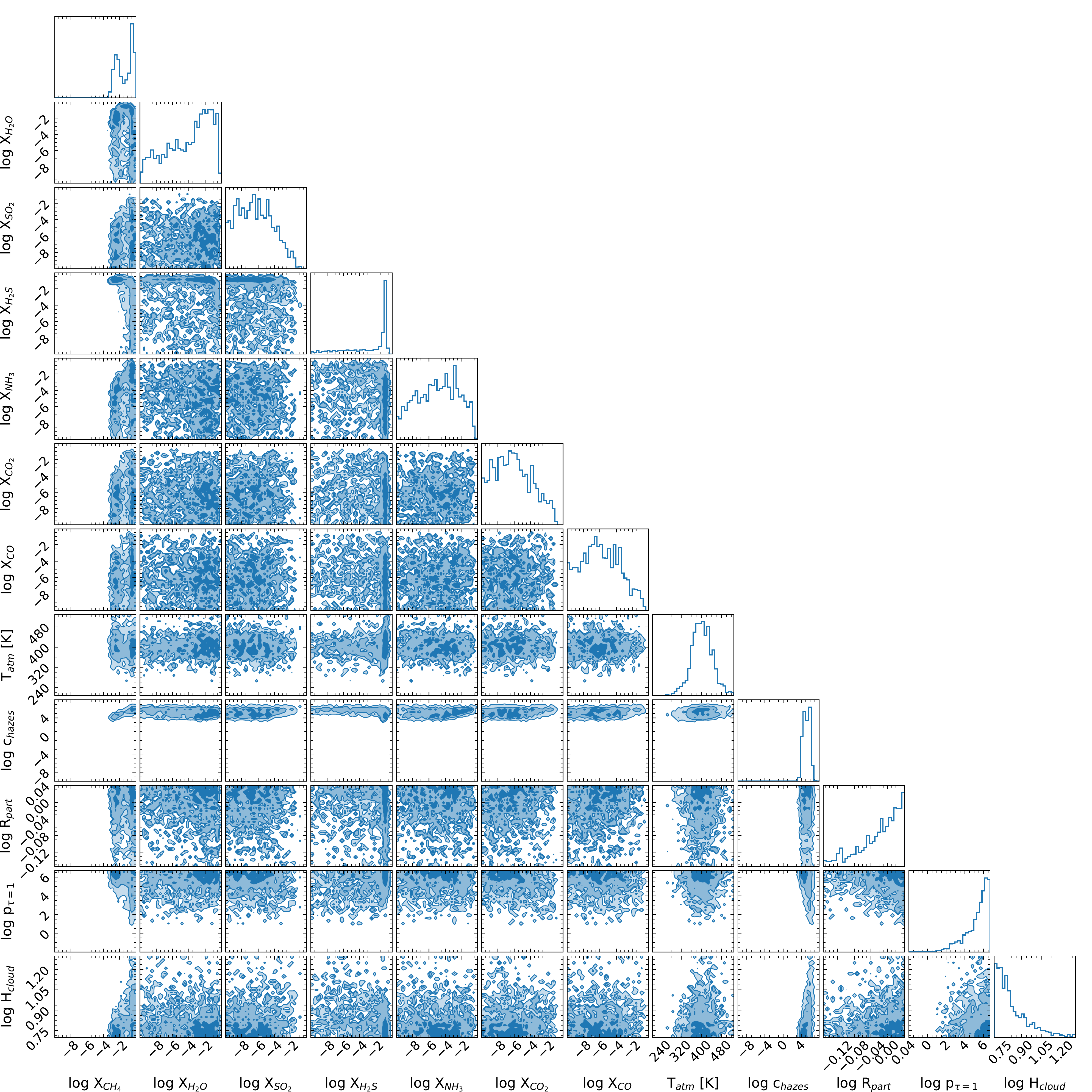}
\end{center}
\caption{\textbf{Parameter inference from the free chemistry retrieval.} Joint and marginalized posterior distributions of the atmosphere parameters obtained from the free chemistry retrieval.  }
\label{fig:wellMixedCorner}
\end{figure*}

\begin{figure*}[t!]
\begin{center}
\includegraphics[width=\linewidth]{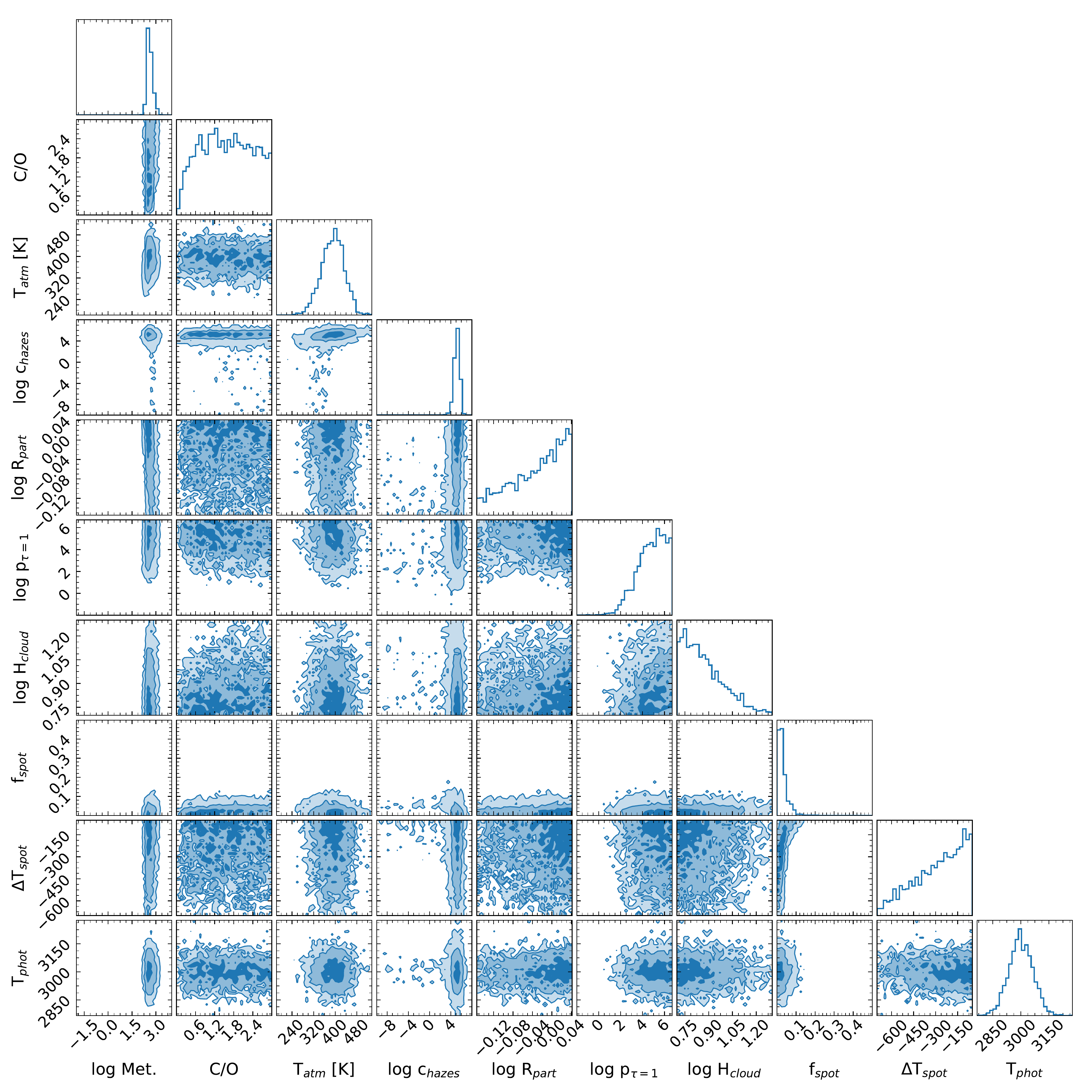}
\end{center}
\caption{\textbf{Parameter inference from the chemically consistent retrieval with stellar contamination.} Joint and marginalized posterior distributions of the atmosphere parameters obtained from the chemically consistent retrieval with stellar spots contamination.  }
\label{fig:ChemEquiCorner}
\end{figure*}

\begin{figure*}[t!]
\begin{center}
\includegraphics[width=0.37\linewidth]{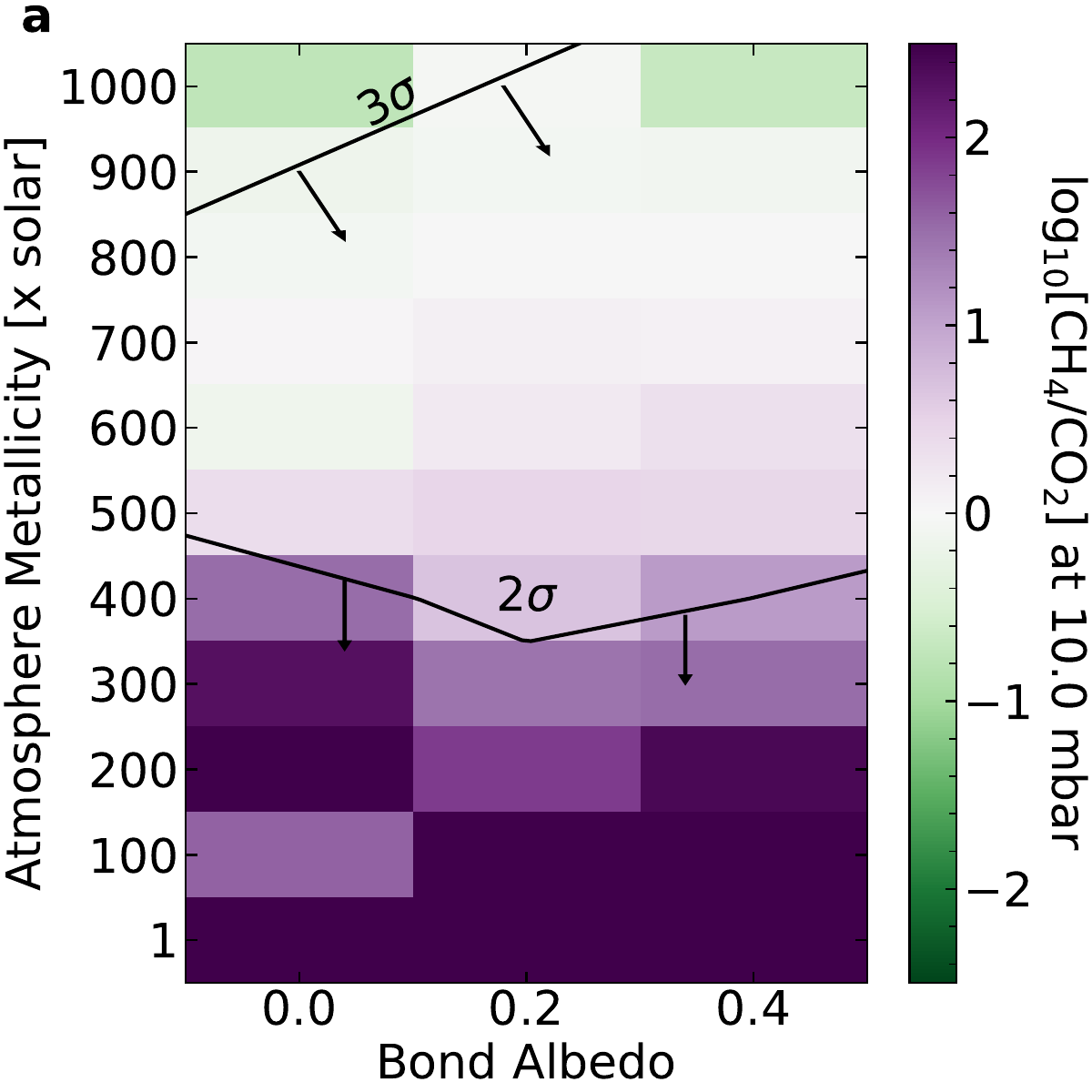}
\includegraphics[width=0.49\linewidth]{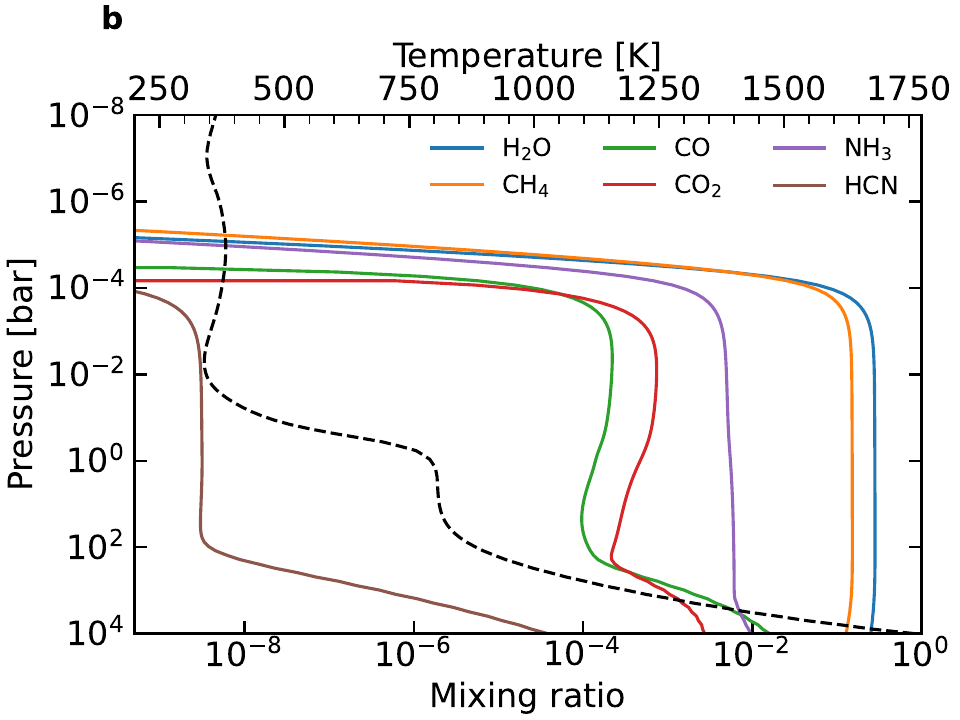}
\end{center}
\caption{\textbf{Grid exploration of the miscible envelope sub-Neptune scenario for LP~791-18\,c.} \textbf{a.} Methane-to-carbon dioxide abundance ratio at 10 mbar for a grid of convective-radiative SCARLET models of LP~791-18\,c for which vertical mixing was modelled using the VULCAN framework with a K$_{zz}$ of 10$^4$ cm$^2$/s. The colorbar is cut at log$_{10}$CH$_4$/CO$_2$ values of +2.5 and -2.5 in order to help visualize the transition from methane dominated to carbon dioxide dominated regimes. The 2- and 3$\sigma$ lower bounds on the methane-to-carbon dioxide abundance ratio derived from the free chemistry retrieval are shown as the black line and arrows. 
\textbf{b.} Mixing ratios of a sample model of the grid shown on the left for the case of 0.0 Bond albedo and 300 $\times$ solar metallicity. The temperature-pressure profile is shown as the black dashed line with the corresponding temperature axis at the top. At the temperature regime of LP~791-18\,c, CH$_4$ strongly dominates over CO$_2$ unless the metallicity is increased above 500 $\times$ solar. }
\label{fig:miscibleEnv}
\end{figure*}

\begin{table*}
\caption{\label{tab:mol_constr} Measured atmospheric properties of LP~791-18\,c. We show the inferred quantities from multiple retrievals for each chemical setup (free or consistent). TLS parameters are also quoted when applicable. The presented uncertainties are the 68\% (1$\sigma$) confidence intervals for all two-sided constraints, and the 95\% (2$\sigma$) upper or lower limits otherwise.}
\centering
\begin{tabular*}{\textwidth}{p{0.35\textwidth}p{0.2\textwidth}p{0.2\textwidth}p{0.2\textwidth}}
\hline
\hline
\textbf{Free Chemistry} &  \\
Parameter & Standard & + TLS & CLR\\
 & retrieval &  & \\
\hline
\textit{Detection} \\
$\log X_{\mathrm{CH_4}}$ & $-1.24^{+0.89}_{-1.34}$ & $-0.49 ^{+0.28} _{-1.54}$ & $>-0.72$\\ 
log c$_\mathrm{Haze}$ & $5.16 ^{+0.75} _{-0.80}$ & $5.09 ^{+0.51} _{-0.69}$ & $5.64^{+0.55}_{-0.51}$\\
\hline
\textit{Other constraints} \\
$\log X_{\mathrm{H_2O}}$ & $< -0.59$ & $< -0.78$ & $< -0.57$\\ 
$\log X_{\mathrm{SO_2}}$ & $ < -2.67$ & $< -1.91$ & $< -2.82$\\ 
$\log X_{\mathrm{H2S}}$ & $>-8.58 $ & $>-8.94$ & $>-14.47$\\ 
$\log X_{\mathrm{NH3}}$ & $< -1.14$ & $< -1.15$ & $< -1.00$\\ 
$\log X_{\mathrm{CO_2}}$ & $< -1.97$ & $< -1.50$ & $< -2.21$\\ 
$\log X_{\mathrm{CO}}$ & $< -1.88$ & $< -1.38$ & $< -2.10 $\\  
\hline
\textit{Stellar contamination} \\
T$_\mathrm{phot, star}$ [K] & - & $3003.43 ^{+66.99} _{-66.10}$ & -\\ 
$\Delta$ T$_\mathrm{spot}$ [K] & - & $>-615.96$ & -\\ 
f$_\mathrm{spot}$ [K] & - & $< 0.07$ & -\\ 
\hline
\textit{Derived quantities} \\
MMW [amu] & $7.68 ^{+2.24} _{-1.48}$ & $9.58 ^{+2.67} _{-2.18}$ & $16.02 ^{+0.18} _{-6.79}$\\
$\log X_{\mathrm{CH_4}}/X_{\mathrm{CO_2}}$ & $>1.07$ & $>0.77$ & $>2.13$\\ 
\hline
\hline
\end{tabular*}
\begin{tabular*}{\textwidth}{p{0.35\textwidth}p{0.3\textwidth}p{0.3\textwidth}} 
\textbf{Chemically consistent}  \\
Parameter & Chemical & +TLS \\
 & equilibrium &  \\
\hline
\textit{Detection} \\
Metallicity [$\times$ solar] & $2.50 ^{+0.10} _{-0.11}$ & $2.61 ^{+0.19} _{-0.13}$ \\ 
log c$_\mathrm{Haze}$ & $5.51 ^{+0.44} _{-0.38}$ & $5.15 ^{+0.46} _{-0.51}$\\
\hline
\textit{Other constraints} \\
C/O & $>0.16$ & $>0.19$ \\ 
\hline
\textit{Stellar contamination} \\
T$_\mathrm{phot, star}$ [K] & - & $3005.05 ^{+59.59} _{-69.69}$ \\ 
$\Delta$ T$_\mathrm{spot}$ [K] & - &  $>-469.94$  \\ 
f$_\mathrm{spot}$ [K] & - & $< 0.06$ \\ 
\hline
\textit{Derived quantities} \\
MMW [amu] & $9.48 ^{+2.17} _{-1.66}$ & $12.07 ^{+1.42} _{-3.09}$ \\
\hline
\hline
\end{tabular*}
\end{table*}


\begin{table*}
\caption{Bayesian model comparison results from our SCARLET atmosphere retrievals in the free chemistry settings. We find strong evidence for the presence of methane, and hazes.}
\centering
\begin{tabular}{lccccc}
\hline
\hline
Retrieval model &  Evidence & Bayes Factor &  N$_\sigma$ & Interpretation\\
                &  ln(Z$_i$)    & B = Z$_{\mathrm{ref}}$/Z$_i$& \\ 
\hline
\textbf{Free Chemistry retrievals} \\
\hline 
All mols. & -2869.97 & Ref. & Ref. \\
CH$_4$ removed & -2878.25 & 3964.89 & 4.47 & Strong detection\\
H$_2$O removed & -2870.19 & 1.25 & 1.40 \\
SO$_2$ removed & -2869.69 & 0.75 & 0.90 \\
H$_2$S removed & -2870.55 & 1.78 & 1.73 \\
NH$_3$ removed & -2870.35 & 1.47 & 1.57 \\
CO$_2$ removed & -2870.63 & 1.93 & 1.80 \\
CO removed & -2870.75 & 2.19 & 1.88 \\
Hazes removed & -2882.46 & 266504.97 & 5.36 & Strong detection\\
\hline
\textit{Hazes vs TLS}\\
with Spots, with Hazes & -2872.33 & Ref. & Ref. \\
with Spots, Hazes removed & -2877.11 & 119.34 & 3.54 & Strong detection\\
\hline
\hline
\end{tabular}
\label{tab:detect2025} 
\end{table*}

\clearpage

\setcounter{figure}{0}
\setcounter{table}{0}
\renewcommand{\figurename}{Supplementary Figure}
\renewcommand{\tablename}{Supplementary Table}

\end{supplementary}

\typeout{}
\clearpage
\bibliography{references.bib, pythonpackages.bib}
\newpage

\clearpage

\end{document}